\documentclass[superscriptaddress,aps,prl,reprint,twocolumn,amsmath,amssymb]{revtex4-2}
\usepackage{graphicx}
\usepackage{hyperref}
\usepackage{amssymb}
\usepackage{slashed}
\usepackage{dcolumn}
\usepackage{amsmath}
\usepackage{bm}
\usepackage{colordvi}
\usepackage{algorithm}
\usepackage{algpseudocode}
\usepackage{multirow}
\usepackage{titlesec}
\usepackage{newtxtext,newtxmath}
\usepackage{dsfont}
\usepackage{xcolor}
\usepackage{mathbbol}

\usepackage{hyperref}
\hypersetup{
    colorlinks=true,
    linkcolor=blue,
    citecolor=blue,     
    urlcolor=blue,
}

\allowdisplaybreaks

\usepackage{mathrsfs}
\makeatletter

\newcommand{\Rmnum}[1]{\expandafter\@slowromancap\romannumeral #1@}
\makeatother

\begin{document}

\title{A composite electron-lattice order: electronic nematicity of 2DEG and polarization density waves at a near-ferroelectric interface}

\author{Fei Yang}
\email{fzy5099@psu.edu}

\affiliation{Department of Materials Science and Engineering and Materials Research Institute, The Pennsylvania State University, University Park, PA 16802, USA}

\author{Zhi-Yang Wang}

\affiliation{Department of Materials Science and Engineering and Materials Research Institute, The Pennsylvania State University, University Park, PA 16802, USA}

\author{Long-Qing Chen}
\email{lqc3@psu.edu}

\affiliation{Department of Materials Science and Engineering and Materials Research Institute, The Pennsylvania State University, University Park, PA 16802, USA}

\date{\today}

\begin{abstract} 
We consider a two-dimensional electron gas (2DEG) formed at a near-ferroelectric interface and strongly coupled to polar phonons. Through a self-consistent microscopic many-body calculation, we show that the coupled system stabilizes a composite electron-lattice ordered state in which the lattice polarization spontaneously forms a polarization density wave (PDW), accompanied by an electronic stripe order in the 2DEG. This intertwined order partially reconstructs the electronic spectrum and generates a twofold quasiparticle anisotropy, giving rise to electronic nematicity at the single-particle level. However, under strong external electric fields, the nematic response becomes dominated by the collective sliding dynamics of the composite order: the sliding motion overwhelms the quasiparticle anisotropy and produces a strongly enhanced nematic signal with higher-order angular harmonics. The theory offers a natural explanation for several anomalous transport and anisotropic responses recently observed at the KTaO$_3$ (111) interface. We also estimate the mean-field transition temperature of this emergent ordered state, obtaining good agreement with experiments, and analyze its evolution with several tuning parameters. The proposed composite order, along with the field-induced crossover from quasiparticle-driven to sliding-dominated nematicity, provides a distinct mechanism of nematicity arising from many-body effects and collective dynamics in critical electron-boson systems, with applicability beyond ferroelectric platforms.
  
\end{abstract}

\maketitle  

{\sl Introduction.---}Owing to their reduced dimensionality, two-dimensional electron gases (2DEGs) formed at oxide interfaces often host emergent electronic phases with profound consequences for transport and collective ordering phenomena that have no direct counterparts in the bulk~\cite{Hwang2012OxideInterfaces,Nakagawa2006PolarCatastrophe,Zubko2011InterfacePhysics,Garcia2010FerroelectricControl}.  Recent experiments on KTaO$_3$-based interfaces~\cite{Hua2024,PhysRevX.15.021018,Liu2021KTaO3_111_SC,Arnault2023KTO111_anisoSC,Zhang2023KTO111_rotSymBreak,Chen2024LAO_KTO_orientation,Chen2021LAO_KTO111_gateSC,PhysRevLett.126.026802,PhysRevX.15.011037,PhysRevX.15.011006,1848-9r5d} have reported a variety of unconventional transport behaviors of 2DEG, including the spontaneous onset of electronic nematicity that develops below a well-defined critical temperature~\cite{PhysRevX.15.021018,Liu2021KTaO3_111_SC,Arnault2023KTO111_anisoSC,Zhang2023KTO111_rotSymBreak,Hua2024}. Moreover, these systems simultaneously develop interfacial polarization~\cite{Dong2026LAO_KTO111_FE_SC,Zhang2025LAO_KTO_FE_SC}, despite the host material remaining paraelectric down to the lowest temperatures in the bulk~\cite{PhysRevB.53.176}. As a result, electronic transport becomes bistable, manifested, for example, in a hysteretic dependence on external electrostatic gating~\cite{Dong2026LAO_KTO111_FE_SC,Zhang2025LAO_KTO_FE_SC}.  Interestingly, as the temperature is further lowered into the superconducting regime, experiments have reported seemingly disparate behaviors regarding the fate of the nematic order. While some studies suggest that nematicity and superconductivity do not coexist~\cite{Liu2021KTaO3_111_SC}, others find that a pre-existing nematic state persists into the superconducting phase and induces a pronounced twofold anisotropy in the superconducting response~\cite{Zhang2023KTO111_rotSymBreak,PhysRevX.15.021018,Arnault2023KTO111_anisoSC,Hua2024}, such that superconductivity is  effectively confined along a preferred in-plane direction. 

 Electronic nematicity, defined as the spontaneous rotational-symmetry  breaking of the  electron  phase~\cite{Kasahara2012BaFe2AsP_nematic,Kivelson1998ElectronicLiquidCrystal,Fradkin2010NematicReview}, typically emerges in strongly correlated systems, such as cuprate~\cite{Yamase2021NematicReview,Lawler2010Cuprate_intraUnitNematic,Hinkov2008YBCO_nematic,PhysRevLett.88.137005,Daou2010YBCO_pseudogapNematic} and iron-based superconductors~\cite{Wang2011RbFeSe_spinwave,Yi2011BaFeCo_orbitalNematic,Chuang2010FeSC_nematicSTM,Chu2010FeSC_resistivityNematic,PhysRevB.81.184508,PhysRevLett.113.237001}, as well as in quantum Hall systems~\cite{PhysRevLett.121.147601}. In most theoretical descriptions, this behavior is attributed either to explicit anisotropies in the underlying electronic band structure~\cite{Fradkin2010NematicReview} or to interaction-driven Pomeranchuk instabilities of the Fermi surface~\cite{PhysRevLett.121.147601,PhysRevB.95.075109,Fradkin2010NematicReview}.  However, purely electronic mechanisms appear insufficient to account for the nematicity phenomenology observed at KTaO$_3$ interfaces. These systems belong to the class of displacive quantum paraelectrics~\cite{PhysRevB.19.3593,Rowley2014FE_QCP,Fujishita2016QuantumParaelectric,yang2024first,PhysRevB.106.224102,PhysRevLett.130.126902,74d5-4hsw,PhysRevMaterials.7.L030801,Li2019THz_SrTiO3_FE}, where a crystal-instability analysis would predict a low-temperature transition to a ferroelectric phase but quantum zero-point fluctuations suppress the formation of  ferroelectric order, leaving the lattice poised at incipient criticality.  Interfaces proximate to lattice quantum criticality, where exceptionally soft collective modes are present~\cite{Rowley2014FE_QCP,yang2024first,PhysRevMaterials.7.L030801,PhysRevB.106.224102}, therefore constitute a  unique platform in which itinerant fermions coexist with nearly critical bosonic modes in the form of soft polar phonons. The system is in principle  susceptible to couplings between electronic degrees of freedom and lattice dynamics, analogous to lattice-driven charge-density-wave (CDW) instabilities~\cite{gruner1988dynamics,gruner1985charge,lee1974conductivity}, and places the electron-lattice system at the brink of a many-body instability even prior to the onset of superconductivity, suggesting that collective order may already emerge in the normal state and drive unconventional symmetry breaking.  

In this work, based on a self-consistent
microscopic many-body theory, we propose that coupling between itinerant electrons and critical polar phonons at a near-ferroelectric interface generically drives the formation of a composite electron-lattice ordered state, in which a lattice polarization density wave (PDW) and an electronic stripe order emerge spontaneously. This intertwined phase reconstructs the electronic spectrum and producing a twofold anisotropy in quasiparticle properties, which  naturally gives rise to electronic nematicity without invoking fine-tuned band anisotropies or purely electronic instabilities. Notably, we show that the nematic response of this composite ordered state is not solely governed by static quasiparticle anisotropy: the dominant contribution under strong external electric fields instead originates from the collective sliding dynamics of the PDW-stripe composite. This sliding motion results in a dramatic enhancement of the nematic response, and naturally generates higher-order angular harmonics beyond simple twofold anisotropy, in quantitative  agreement with recent experimental observations  at the KTaO$_3$ (111) interface~\cite{PhysRevX.15.021018,Zhang2023KTO111_rotSymBreak}.  We also estimate the transition temperature of this composite electron-lattice order, finding good agreement with experimentally observed onset temperatures~\cite{PhysRevX.15.021018}, and identify several tunable parameters that provide clear routes for independent verification. The composite electronic-lattice ordering offers a generic understanding of nematicity at near-ferroelectric interfaces beyond KTaO$_3$, and suggests broad applicability to low-dimensional systems characterized by strong coupling between electrons and critical bosonic modes beyond ferroelectric platforms.

{\sl Model.---}We consider a 2DEG coupled to a polar optical phonon mode at an interface. In polar materials, the longitudinal optical (LO) mode is strongly hardened by long-range Coulomb interactions~\cite{PhysRevLett.131.046801,PhysRevLett.72.3618,PhysRevB.81.024102,yang2025ferroelectric}. We therefore focus on the low-energy transverse optical (TO) mode, which controls the proximity to ferroelectricity. As the electronic motion is confined to the interface plane, only the TO displacement polarized perpendicular to the interface (along the $z$ direction) couples efficiently to the 2DEG~\cite{norman2026superconductivity,Liu2023KTO_tunableSC,yang2025ferroelectric}. Retaining this relevant degree of freedom (see Sec.~SI), the Hamiltonian reads~\cite{abrikosov2012methods,mahan2013many} 
\begin{eqnarray}
&&\!\!\!\!\!\!H=\!\sum_{{\bf k}\alpha}\psi_{{\bf k}\alpha}^{\dagger}\xi_{{\bf k},\alpha}\psi_{{\bf k}\alpha}\!+\!\frac{1}{2}\!\int\!\!{{d{\bf r}}{d{\bf r'}}}\phi_{z}({\bf r}){S({{\bf r}\!-\!{\bf r'}})}\phi_{z}({\bf r}')\nonumber\\
&&\!\!\!\!\!+\!\!\int\!\!{d{\bf r}}\Big[\frac{a}{2}\phi^2_{\rm z}({\bf r})\!+\!\frac{b}{4}\phi^4_{\rm z}({\bf r})\Big]\!+\!\!\sum_{{\bf qk}\alpha}\,g^{\alpha}_{{\bf k+{\bf q},k}}\psi^{\dagger}_{{\bf k+q}\alpha}\psi_{{\bf k}\alpha}\phi_z({\bf q}),~~~~
\end{eqnarray}
where $\psi^{\dagger}_{{\bf k}\alpha}$ and $\psi_{{\bf k}\alpha}$ are fermionic creation and annihilation operators with momentum ${\bf k}$ in band $\alpha$, and $\xi_{{\bf k},\alpha}$ is the  dispersion measured from the chemical potential $\mu$.  The bosonic field is defined as $\phi_z=u_z\sqrt{M}$, with $u_z$ the ionic displacement along the polar axis and $M$ the effective ionic mass. 
The parameter $a$ represents the harmonic restoring coefficient
of the TO mode, while $b$ characterizes the local quartic
anharmonicity.
The nonlocal kernel $S({{\bf r}-{\bf r'}})$ generates the phonon dispersion~\cite{PhysRevLett.131.046801,PhysRevLett.72.3618,PhysRevB.81.024102,yang2025ferroelectric}; taking its Fourier transform as $S({\bf q})=v^2 q^2$
yields the spectrum $\omega_{\bf q}^2 = a + v^2 q^2$. The phonon sector follows from expanding the lattice potential
energy around the centrosymmetric equilibrium configuration~\cite{kittel1963quantum}. The parameter $a$ controls the proximity to the ferroelectric instability: for $a>0$ the system remains centrosymmetric (paraelectric), whereas $a\to0^+$ signals critical softening of the TO mode.

The quantity $g^{\alpha}_{{\bf k+{\bf q},k}}$ denotes the effective electron–phonon coupling vertex. In a centrosymmetric bulk system, the TO displacement does not couple linearly to the electronic density, and the leading coupling vanishes~\cite{PhysRevB.100.226501}. At an interface, however, inversion symmetry is explicitly broken, allowing a finite linear coupling to the TO mode.  Two microscopic mechanisms can generate such a vertex (see Sec.~SII for details).
First, interfacial polar displacements produce bound polarization charges and modify the electrostatic potential seen by the 2DEG~\cite{Hwang2012OxideInterfaces,Nakagawa2006PolarCatastrophe,Zubko2011InterfacePhysics,Garcia2010FerroelectricControl}. This yields a density-displacement coupling 
$g^{\alpha}_{{\bf k+{\bf q},k}}\propto e\,F({\bf q})$, where $F({\bf q})$ is an interface form factor determined by the spatial distribution of polarization charge and screening.     Second, interfacial inversion breaking  induces Rashba spin-orbit coupling. Then, expanding  to linear order in $u_z$ produces an effective intraband vertex in the helicity basis ($\nu=\pm$), $g^{\nu}_{{\bf k+{\bf q},k}}
\propto \nu|{\bf k+q/2}|\delta\lambda_{R}$,
where 
$\delta\lambda_{R}$ denotes the modulation
of the Rashba coupling by lattice vibrations~\cite{PhysRevB.105.224503,norman2026superconductivity,PhysRevResearch.5.023177,7l88-12m5}. 

Quantizing the phonon field as $\phi_z({\bf q})=
(b_{\bf q}+b^{\dagger}_{-{\bf q}})/({\sqrt{2\omega_{\bf q}}})$,  
 the interface-enabled linear coupling renders the interaction formally analogous to a Fr\"ohlich Hamiltonian~\cite{frohlich1954theory}, albeit involving a soft TO mode instead of a bulk LO phonon. Following the approach of Lee, Rice, and Anderson for 1D Fr\"ohlich-type coupling~\cite{lee1974conductivity}, generalized here to 2D, one can single out the polar phonon mode at ${\bf Q}$ and introduce an order parameter 
\begin{equation}\label{OP}
\Delta = -2 g\, \langle \phi_z({\bf Q}) \rangle, \quad  \Delta^* =- 2 g\, \langle \phi_z(-{\bf Q}) \rangle,
\end{equation}
where 
$g^{\alpha}_{{\bf k+Q},{\bf k}}$ is approximated by a constant 
$g$ within the low-energy sector (see Sec.~III for details), evaluated near Fermi surface.  The resulting mean-field Hamiltonian reads
\begin{eqnarray}
{\bar H}=\frac{1}{2}\sum_{\bf k,\alpha}\Psi^{\dagger}_{\bf k\alpha}\left(\begin{array}{cc}
\xi_{{\bf k_+},\alpha} & -\Delta \\
-\Delta^*& \xi_{{\bf k_-},\alpha} 
\end{array}\right)\Psi_{\bf k\alpha}\!+\!\frac{\omega^2_{\bf Q}|\Delta|^2}{4g^2}\!+\!\frac{3b|\Delta|^4}{32g^4},
\end{eqnarray}
with the spinor $\Psi^{\dagger}_{\bf k\alpha}=(\psi^{\dagger}_{{\bf k_+},\alpha},\psi^{\dagger}_{{\bf k_-},\alpha})$ and ${\bf k}_{\pm}={\bf k}\pm{\bf Q}/2$. 

Diagonalizing this BCS-like mean-field Hamiltonian~\cite{abrikosov2012methods,mahan2013many,schrieffer1964theory}, we
obtain the quasiparticle spectrum
\begin{equation}
E_{{\bf k}\alpha}^{\pm}=({\xi_{\bf k_-,\alpha}+\xi_{\bf k_+,\alpha}})/{2}\pm{E_{\bf k\alpha}},
\end{equation}
and the self-consistency mean-field gap equation
\begin{equation}
\frac{\omega^2_{\bf Q}}{g^2}+\frac{3b|\Delta|^2}{4g^4}=\sum_{{\bf k}\alpha}\frac{f(E_{\bf k\alpha}^-)-f(E_{\bf k\alpha}^+)}{2E_{\bf k\alpha}},
\end{equation}
with $E_{\bf k\alpha}=\sqrt{[(\xi_{\bf k_-,\alpha}-\xi_{\bf k_+,\alpha})/2]^2+|\Delta|^2}$ and $f(x)$ being the Fermi distribution function. In the absence of the anharmonic term ($b=0$), the self-consistency equation above formally resembles a BCS-like gap equation in the particle-hole channel, and suggests an effective interaction of order ${\sim}g^2/\omega^2_{\bf Q}$. 

If this self-consistent gap equation admits a nontrivial solution, it implies that the initial uniform paraelectric state with itinerant electrons at the interface becomes unstable toward the formation of a modulated order. Notably, near the ferroelectric critical point ($a=0^+$), the divergence of the polar susceptibility amplifies the effective interaction. As a consequence, a finite-${\bf Q}$ stripe instability may develop even in the absence of perfect Fermi-surface nesting. In this sense, the critical softening of the polar mode compensates for imperfect nesting by reducing bosonic stiffness that penalizes density modulation. This  nearly critical polar mode, owing to its strong coupling to electrons, strongly hybridizes distinct regions of the Fermi surface, such that the system becomes energetically predisposed toward the formation of a modulated state (see Sec.~SV.~A).

To make the above mechanism more transparent, we approximate the electronic dispersion by parabolic bands, $\xi_{{\bf k},\alpha}={k^2}/(2m)-\mu$, excluding any Fermi-surface nesting effects in the 2DEG. We consider the KTaO$_3$ (111) interface, whose low-energy sector consists of three nearly degenerate conduction bands derived from the $t_{2g}$ manifold~\cite{norman2026superconductivity,Liu2023KTO_tunableSC,7l88-12m5}. Using an experimentally estimated electron-phonon coupling constant  $\lambda\approx0.26$~\cite{norman2026superconductivity,Liu2023KTO_tunableSC}, an effective mass $m\approx0.34m_e$~\cite{Zou2015LTO_KTO_2DEG,PhysRevLett.108.117602} and electron density $n_e=6\times10^{13}/$cm$^2$, we evaluate the self-consistent gap equation. The remaining model parameters of polar phonons are specified in the Supplementary Materials (Sec.~SVI) and are taken from independent  measurements. Notably, $|{\bf Q}|$ in the simulation is determined self-consistently from the maximum of $|\Delta({\bf Q})|$, corresponding to the free-energy minimum. 
Under realistic material parameters, we consistently obtain an optimal ordering wavevector $|{\bf Q}|\approx2k_F$, 
despite the absence of Fermi-surface nesting in this 2DEG. Physically, by analogy with the standard BCS analysis~\cite{fulde1964superconductivity,yang2018fulde,yang21theory,PhysRevB.95.075304,PhysRevB.98.094507}, a stable nontrivial solution of the gap equation requires the band-center (effective Zeeman-field) term 
$\xi_{{\bf k}_-,\alpha} + \xi_{{\bf k}_+,\alpha}$ to be small, 
so that the hybridized states remain close to the Fermi level. 
For a parabolic dispersion, this condition leads to $Q^2/(8m) \approx \mu$, 
which naturally gives $|{\bf Q}| \approx 2k_F$. 
The resulting $2k_F$ scale here thus reflects a phase-space matching condition rather than a nesting-driven instability, while this resembles the characteristic $2k_F$ scale in conventional CDW  systems~\cite{gruner1988dynamics,gruner1985charge,lee1974conductivity}.

{\sl Density-wave modulation.---}We first elucidate the real-space structure associated with the order parameter $\Delta$. The macroscopic polarization along the interface normal is defined as $P_z=z^*\langle{u_z}\rangle/\Omega_{\rm cell}=\frac{z^*}{\sqrt{M}\Omega_{\rm cell}}\langle\phi_z({\bf r})\rangle$ with $z^*$ and $\Omega_{\rm cell}$~\cite{yang2024first,cochran1981soft,cochran1961crystal,cowley1965theory,cochran1969dynamical,yelon1971neutron,tadmor2002polarization,tadmor2002polarization,zhong1994phase,PhysRevB.52.6301,PhysRevB.55.6161,PhysRevLett.84.5427,naumov2004unusual,PhysRevLett.97.157601,PhysRevLett.108.257601}, where $z^*$ is the Born effective charge~\cite{tadmor2002polarization,tadmor2002polarization,zhong1994phase,PhysRevB.52.6301,PhysRevB.55.6161,PhysRevLett.84.5427,naumov2004unusual,PhysRevLett.97.157601,PhysRevLett.108.257601} and  $\Omega_{\rm cell}$ is the unit-cell volume. By definition in Eq.~(\ref{OP}), a finite $|\Delta|$ directly implies condensation of the transverse optical mode at wavevector ${\bf Q}$, i.e., $\langle\phi_z({\bf r})\rangle=\langle\phi_z({\bf Q})\rangle{e^{i{\bf Q}\cdot{\bf r}}}+c.c$, and hence, this leads to a spatially modulated polarization
\begin{eqnarray}
P_z({\bf r})=-{z^*}/{(g\sqrt{M}\Omega_{\rm cell})}|\Delta|\cos({\bf Q}\cdot{\bf r}+\Theta).
\end{eqnarray}
Meanwhile, the electronic density $\rho_e({\bf r})
=\sum_{\bf q}\langle\rho_e({\bf q})\rangle\,e^{i{\bf q}\cdot{\bf r}}
\!+\! \text{c.c.}$, with density operator $\rho_e(-{\bf q})=\sum_{\bf k\alpha}
\psi^{\dagger}_{{\bf k}+{\bf q/2},\alpha}
\psi_{{\bf k-q/2},\alpha}$, develops a finite ${\bf Q}$-component (density-wave modulation) 
\begin{align}
\rho_e({\bf r})=2\chi|\Delta|
\cos\!\left({\bf Q}\cdot{\bf r}+\Theta\right),
\end{align}
where {\small $\chi={\omega^2_{\bf Q}}/(2g^2)+{b|\Delta|^2}/(8g^4)$}. Here we have parametrized the order parameter as $\Delta=|\Delta|e^{i\Theta}$, where 
 $\Theta$ is the phase associated with translational symmetry breaking, which directly determines the spatial shift of the modulation in real space.

\begin{figure}[H]
  {\includegraphics[width=8.2cm]{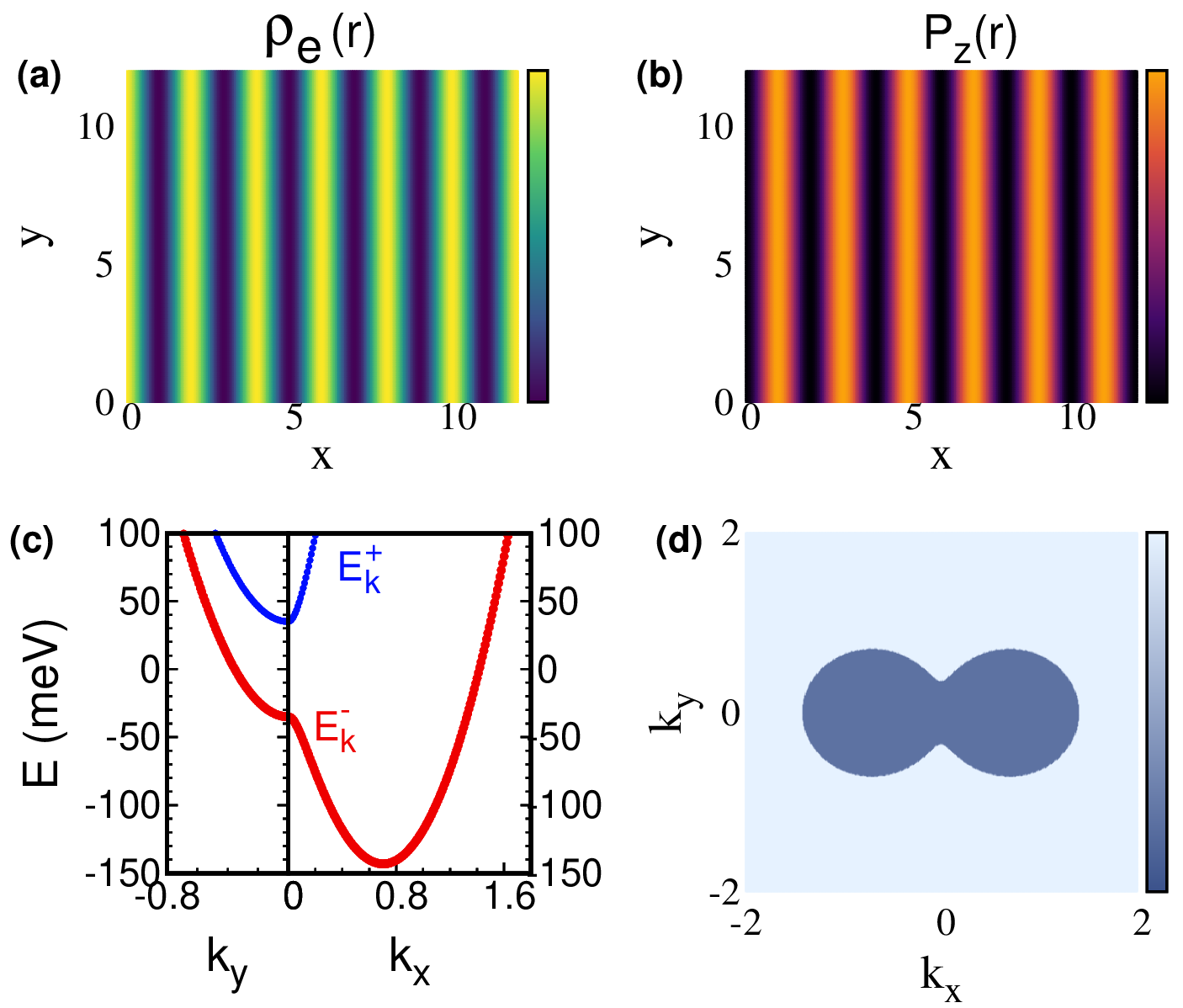}}
\caption{Formation of ({\bf a}) an electronic density-wave modulation with stripe order and ({\bf b}) a PDW.
({\bf c}) Corresponding quasiparticle spectra and ({\bf d}) the associated Fermi surface.
The $x$ axis is chosen parallel to ${\bf Q}$, whose spontaneous selection signals rotational symmetry breaking. The calculated value of $|\Delta(T=0)|$ is $0.35$ meV ($\ll E_F$); it is  rescaled by a factor of 100 for schematic purposes in ({\bf c}) and ({\bf d}).}
\label{figyc}
\end{figure}

Consequently, a finite amplitude $|\Delta|$ corresponds to a phase-locked real-space modulation of both lattice polarization and electronic density (Sec.~SV.~A). The ordered phase is thus a composite electron-lattice density-wave state, characterized by the simultaneous formation of an electronic stripe order in the 2DEG [as shown Fig.~\ref{figyc}(a)] and a PDW [as shown in Fig.~\ref{figyc}(b)].

{\sl Electronic nematicity.---}
The resulting quasiparticle dispersion and reconstructed Fermi surfaces are shown in Fig.~\ref{figyc}(c) and (d). The momentum ${\bf k}$ is defined in the reduced-zone representation associated with the ordering wave vector ${\bf Q}$, and the $x$ axis is chosen parallel to ${\bf Q}$, whose spontaneous selection signals rotational symmetry breaking. For $|\Delta|=0$, the quasiparticle energies reduce to the two dispersions $\xi_{{\bf k}\pm{\bf Q}/2}$, corresponding to parabolic branches shifted by $\pm{\bf Q}/2$. 
In this case, $\xi_{{\bf k}\pm{\bf Q}/2}$ remain degenerate along the $k_y$ direction, and the associated Fermi surface consists of two touching circular pockets. 
These pockets arise solely from the finite-${\bf Q}$ representation of the dispersion and do not reflect any intrinsic distortion of the underlying Fermi surface, which remains rotationally symmetric in the normal state.

When a finite $|\Delta|$ develops, states at ${\bf k}\pm{\bf Q}/2$ hybridize, opening avoided crossings at their intersections. The quasiparticle dispersion deviates from the parabolic form, and the degeneracy along the $k_y$ direction is lifted by gap opening at the band crossings, as shown in Fig.~\ref{figyc}(c). This lifting of the $k_y$ degeneracy signals spontaneous breaking of rotational symmetry. The two originally shifted circular Fermi pockets then become reconstructed through hybridization, evolving into characteristic peanut-shaped contour [Fig.~\ref{figyc}(d)] due to gapping along the band-crossing lines. This anisotropic Fermi-surface reconstruction directly manifests electronic nematicity in the quasiparticle spectrum as a consequence of the emergent $|\Delta|$.

\begin{figure}[H]
  {\includegraphics[width=8.4cm]{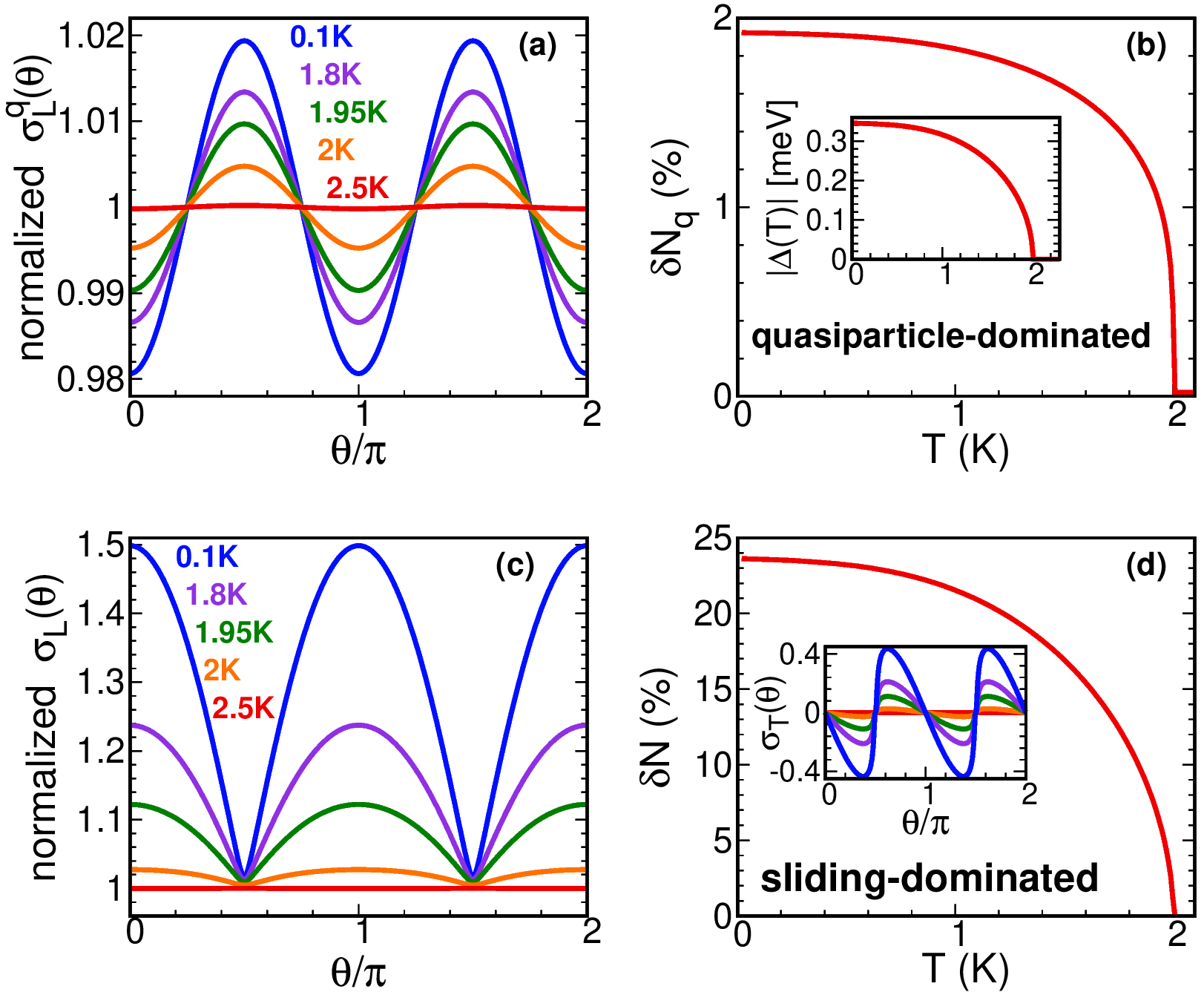}}
\caption{Angular dependence of ({\bf a}) the quasiparticle-dominated conductivity and ({\bf c})  the sliding-motion-dominated conductivity,  at different temperatures. Temperature dependence of the transport nematicity $\delta N$ in ({\bf b}) the quasiparticle-dominated regime and ({\bf d}) the sliding-dominated regime. Inset of ({\bf b}): temperature dependence of the order parameter. Inset of ({\bf d}): transverse conductivity at different temperatures. Model parameters are given in Sec.~SVI.}
\label{figyc2}
\end{figure}

At low temperatures, the quasiparticle contribution to the dc conductivity
is directly evaluated within the relaxation-time approximation~\cite{mahan2013many,kittel1963quantum} as (see Sec.~SIV~A)
\begin{equation}
\sigma^{q}_{ij}
=  e^2\tau \int_{\rm FS}\frac{d\ell}{(2\pi)^2}
\frac{v^-_i({\bf k})v^-_j({\bf k})}{|{\bf v}^-({\bf k})|},
\quad i,j=x,y,
\label{eq:sigma_q_FS}
\end{equation}
where ${\bf v}^-({\bf k})=\nabla_{\bf k} E_{\bf k}^-$ denotes the quasiparticle group velocity.
For finite $|\Delta|$, the reconstructed Fermi contour becomes strongly anisotropic (peanut-shaped), resulting in distinct angular dependences of $v_x$ and $v_y$ along the Fermi surface. Consequently, $\sigma^{q}_{xx}\neq \sigma^{q}_{yy}$~\cite{PhysRevX.15.021018}, giving rise to a finite nematic response in quasiparticle transport.  As shown in Fig.~\ref{figyc2}(a), the longitudinal quasiparticle conductivity 
\(\sigma^q_L(\theta)\), defined as the projection of the conductivity tensor along the direction of the applied electric field \(\hat{\bf e}_\theta=(\cos\theta,\sin\theta)\),  
$\sigma^q_L(\theta)
= \hat e_i \sigma^{q}_{ij} \hat e_j
= \sigma^{q}_{xx}\cos^2\theta
+ \sigma^{q}_{yy}\sin^2\theta$, exhibits a characteristic twofold anisotropy of the form
\begin{equation}
\sigma^q_L(\theta)
= {\bar \sigma}
+ [(\sigma^{q}_{xx}-\sigma^{q}_{yy})/2] \cos 2\theta ,
\label{eq:sigma_theta}
\end{equation} 
consistent with $C_2$ symmetry breaking. The quasiparticle nematicity in the transport, defined as $\delta{N}_q=(\sigma^{q}_{xx}-\sigma^{q}_{yy})/(2{\bar \sigma})$, scales with the order parameter magnitude $|\Delta|$. As shown in Fig.~\ref{figyc2}(b), $\delta{N}_q$ follows the same temperature dependence as $|\Delta|$ [inset of Fig.~\ref{figyc2}(b)] and vanishes at the critical temperature where the order parameter disappears.

The quasiparticle contribution dominates only at weak-field (pining) regime. In the strong-field regime $E_x>E_T$ (where $E_x$ is the $x$ component of the applied field ${\bf E}$), the density-wave condensate becomes depinned and acquires a finite collective sliding velocity~\cite{gruner1988dynamics,gruner1985charge}. The resulting collective current is directed along the ordering wavevector ${\bf Q}$ and the corresponding sliding contribution to the conductivity reads (see Sec.~SIV~B for the detailed derivations of the sliding motions) 
\begin{equation}
\sigma_{ij}^s={4e^2\chi|\Delta|}/({\gamma}Q^2)(1\!-\!E_T/|E_x|)e^{-E_0/|E_x|}\delta_{ix}\delta_{jx},
\end{equation}
where $\gamma$ is the damping coefficient, $E_T$ is the threshold field for depinning, $E_0$ the pinning barrier. This contribution vanishes below $E_T$, and increases nonlinearly above threshold. 

The total longitudinal conductivity $\sigma_L(\theta)=\sigma^q_L(\theta)+\sigma^s_L(\theta)$ and transverse  conductivit $\sigma_T(\theta)=\sigma^q_T(\theta)+\sigma^s_T(\theta)$, including both quasiparticle and condensate contributions, as functions of the field angle $\theta$, are plotted in Fig.~\ref{figyc2}(c) and the inset of Fig.~\ref{figyc2}(d), respectively. Because both the sliding channel and its driving field are highly directional, both the longitudinal and transverse conductivities exhibit pronounced higher-order angular harmonics beyond the simple twofold $\cos(2\theta)$ dependence expected from purely band-structure or quasiparticle-driven scenarios. Due to this directional nature, once the sliding channel dominates, it overwhelms the intrinsic quasiparticle anisotropy; compared to the quasiparticle contribution $\delta{N_q(T=0)}$ in Fig.~\ref{figyc2}(b), the onset of collective sliding substantially enhances the nematic signal $\delta{N}=[\sigma_L(0)-\sigma_L(\pi/2)]/(2{\bar \sigma})$ in transport [Fig.~\ref{figyc2}(d)]. 
 
\begin{figure}[H]
  {\includegraphics[width=8.7cm]{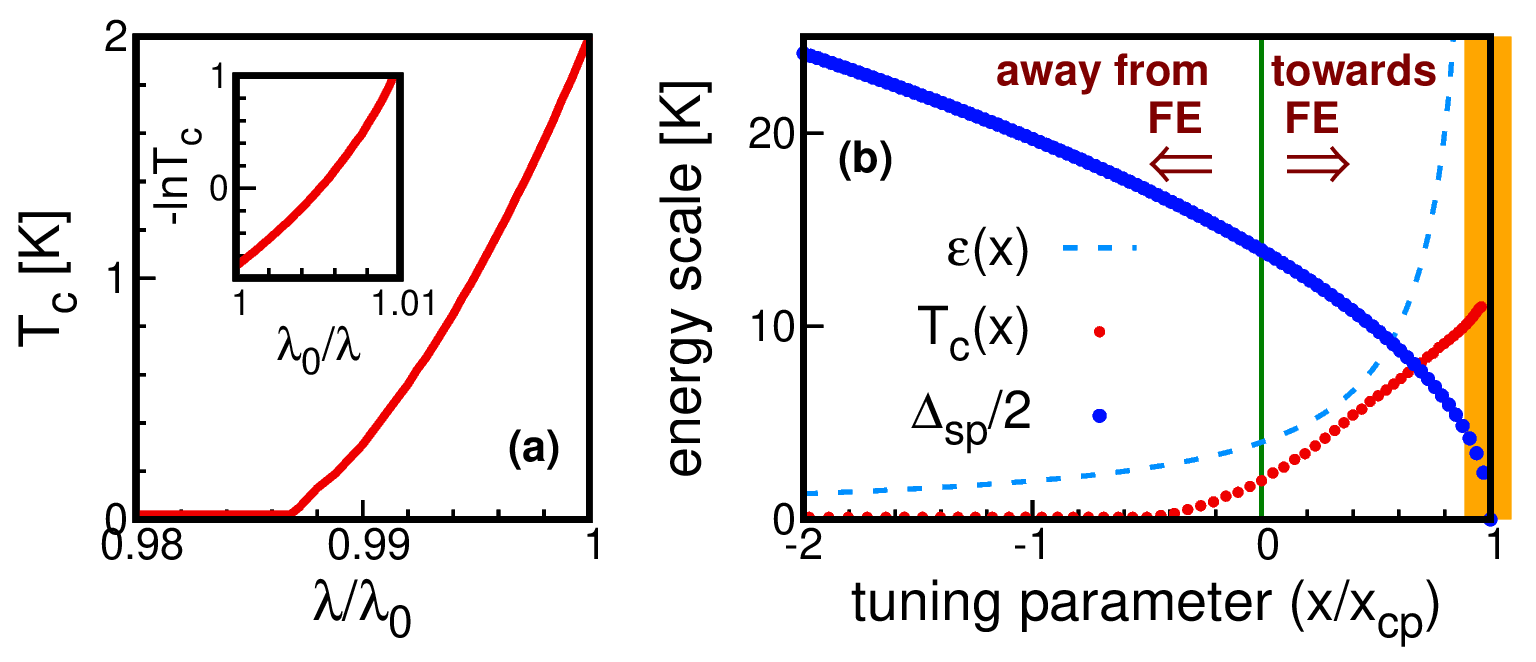}}
 \caption{({\bf a}) Transition temperature $T_c$ as a function of
  electron-phonon coupling constant ${\lambda}$. Inset: $\ln T_c$ as a function of $1/\lambda$, showing a clear deviation from the linear dependence expected in conventional BCS scaling. Here $\lambda_0=0.26$. ({\bf b}) Schematic illustration of $T_c$ and the TO-mode gap
$\Delta_{\rm sp}=\sqrt{a}$ versus the tuning parameter $x/x_{\rm cp}$.
A phenomenological linear softening,
$a(x)=a(0)(1-x/x_{\rm cp})$, is assumed to illustrate the enhancement of $T_c$
toward the ferroelectric critical point.
The dashed curve denotes the dielectric response $\varepsilon(x)$, which grows
and formally diverges at criticality (orange region).} 
\label{figyc3}
\end{figure}

{\sl Transition temperature.---}As illustrated in Fig.~\ref{figyc3}(a), reducing the electron-phonon coupling strength rapidly suppresses the transition temperature $T_c$, suggesting a strong sensitivity of the composite order to the interaction. From the (111) to (110) and (001) interfaces, the effective coupling strength is experimentally realized to decrease~\cite{Liu2023KTO_tunableSC,Liu2021KTaO3_111_SC,Chen2024LAO_KTO_orientation} partially due to the reduced orbital degeneracy of the low-energy manifold and the accompanying changes in the band structure~\cite{norman2026superconductivity}. The composite order is therefore most favorable on the (111) interface, weaker on (110), and most strongly suppressed on (001), consistent with the experimentally observed orientation-dependent nematicity, with a higher nematic transition temperature on the (111) interface~\cite{Zhang2023KTO111_rotSymBreak,PhysRevX.15.021018,Arnault2023KTO111_anisoSC} than on (110)~\cite{Hua2024}. 

Tuning the system closer to or further away from the ferroelectric critical point, via strain~\cite{Li2025}, pressure~\cite{PhysRevB.4.2696}, or  chemical substitution~\cite{PhysRevB.53.5240,PhysRevLett.39.1158,PhysRevLett.96.227602,Rischau2017,PhysRevB.90.165309,PhysRevLett.68.232,He2025,Andrews1985}, provides an additional control knob for transition temperature $T_c$. As shown in Fig.~\ref{figyc3}(d), as the polar mode softens toward criticality via tuning, the $T_c$ is enhanced, reaching a maximum at the critical point. Conversely, tuning away from criticality rapidly weakens the soft-mode enhancement, causing the composite order to collapse, and since the system is already near the ferroelectric critical regime, this suppression implies that the ordered state is in fact stabilized only in a narrow region around criticality.

{\it Discussions.---}The proposed scenario of instability differs from interpretations that attribute the observed anisotropy primarily to anisotropic superfluid stiffness, with the anisotropy assumed to persist into the normal state.  
In contrast, our theory reveal the formation of a distinct composite ordered phase that develops  above the superconducting transition, consistent with understanding in Refs.~\cite{Liu2021KTaO3_111_SC,PhysRevX.15.021018}. While some studies have suggested that electronic nematicity or stripe order requires magnetic proximity from a neighboring material (e.g., EuO)~\cite{PhysRevB.103.035115,Hua2024}, the transport nematicity has also been  observed in LaAlO$_3$/KTaO$_3$~\cite{PhysRevX.15.021018} and (a)-YAlO$_3$/KTaO$_3$~\cite{Zhang2023KTO111_rotSymBreak} interfaces, where both LaAlO$_3$ and  (a)-YAlO$_3$ are non-magnetic.  
The proposed composite order with electronic nematicity here points to a generic symmetry-breaking instability of KTaO$_3$-based interfaces, arising intrinsically from the interfacial electron-phonon coupling and the proximity to ferroelectric criticality. It naturally accounts for the large nematic signal in transport measurements~\cite{PhysRevX.15.021018}, yields a transition temperature $T_c\sim2~$K, consistent with the observed onset scale~\cite{PhysRevX.15.021018,Zhang2023KTO111_rotSymBreak,Liu2021KTaO3_111_SC}.  The resulting angular dependence of the longitudinal  [Fig.~\ref{figyc2}(c)] and transverse  (inset of Fig.~\ref{figyc2}(d)) conductivities exhibit clear higher-order angular harmonics beyond the simple twofold response, in agreement with the experimentally observed angular profile~\cite{Zhang2023KTO111_rotSymBreak,PhysRevX.15.021018}. The emergence of interfacial polarization~\cite{Dong2026LAO_KTO111_FE_SC,Zhang2025LAO_KTO_FE_SC}, despite the bulk material remaining paraelectric down to the lowest temperatures~\cite{PhysRevB.53.176}, follows naturally from the PDW  predicted here, and electronic stripe order has also been reported experimentally~\cite{Hua2024}. We therefore conclude that the experimental phenomenology at the KTaO$_3$-based interface is consistent with the emergence of a composite electron-lattice instability. We call for further experimental tests of this scenario, particularly through direct measurements of threshold-field behavior, detection of the polarization modulation wavevector and electronic ordering, exploration of orientation-dependent effects, and systematic tuning using strain, pressure, or carrier doping.

To elucidate the central physics, the present analysis adopts an isotropic parabolic-band approximation, and the symmetry breaking here  reduces the in-plane rotational symmetry from continuous isotropy to $C_2$. In reality,  the KTaO$_3$(111) interface possesses a  threefold lattice symmetry~\cite{PhysRevB.103.035115,norman2026superconductivity,7l88-12m5}. The resulting single-particle band-structure anisotropy is expected to further cooperate with the composite order, reinforcing both its stability and the resulting nematic anisotropy.

{\it Acknowledgments.---}F.Y. acknowledges insightful and valuable discussions with M.R. Norman. This work is supported by the US Department of Energy, Office of Science, Basic Energy Sciences, under Award Number DE-SC0020145 as part of Computational Materials Sciences Program. F.Y. and L.Q.C. appreciate the  generous support from the Donald W. Hamer Foundation through a Hamer Professorship at Penn State.

%
\end{document}


\title{A composite electron-lattice order: electronic nematicity of 2DEG and polarization density waves at a near-ferroelectric interface (Supplementary Materials)}

\author{Fei Yang}
\email{fzy5099@psu.edu}

\affiliation{Department of Materials Science and Engineering and Materials Research Institute, The Pennsylvania State University, University Park, PA 16802, USA}

\author{Zhi-Yang Wang}

\affiliation{Department of Materials Science and Engineering and Materials Research Institute, The Pennsylvania State University, University Park, PA 16802, USA}

\author{Long-Qing Chen}
\email{lqc3@psu.edu}

\affiliation{Department of Materials Science and Engineering and Materials Research Institute, The Pennsylvania State University, University Park, PA 16802, USA}

\maketitle 

\section{Polar-phonon model}

In this section, for completeness, we present the standard optical-phonon Hamiltonian describing the lattice displacement field ${\bf u}^{\rm op}$ in lattice dynamics~\cite{PhysRevLett.131.046801,PhysRevLett.72.3618,PhysRevB.81.024102,yang2025ferroelectric},
\begin{eqnarray}
H_{\rm op}&=&{\int{d{\bf r}}}\Big[\frac{1}{2}M_{\rm op}(\partial_tu^{\rm op}_{\alpha})^2+\frac{1}{2}u^{\rm op}_{\alpha}({\bf r})\Delta^2_0u^{\rm op}_{\alpha}({\bf r})+\frac{b}{4}\big(u^{\rm op}_{\alpha}({\bf r})\big)^2\big(u^{\rm op}_{\beta}({\bf r})\big)^2\Big]+\frac{1}{2}\int{d{\bf r}}d{\bf r '}u^{\rm op}_{\alpha}({\bf r})S_{\alpha\beta}({\bf r-r'})u^{\rm op}_{\beta}({\bf r'}),
\end{eqnarray}
where $\Delta_0^2$ denotes the local harmonic restoring force of the optical mode, $b$ is the strength of the local quartic self-interaction; $S_{\alpha\beta}({\bf r-r'})$ represents the nonlocal elastic restoring interaction responsible for phonon dispersion, with the Fourier transform, 
 \begin{equation}
  S_{\alpha\beta}({\bf q})=v^2q^2\delta_{\alpha\beta}+h^2q_{\alpha}q_{\beta}/q^2-w^2q_{\alpha}q_{\beta}.
 \end{equation} 
 Here, $h^2q_{\alpha}q_{\beta}/q^2$ represents the long-range dipole–dipole interaction, and $w^2q_{\alpha}q_{\beta}$ accounts for a short-range correction.
 
In the harmonic limit ($b=0$), diagonalization yields the well-known transverse-optical (TO) and longitudinal-optical (LO) phonon modes~\cite{PhysRevLett.131.046801,PhysRevLett.72.3618,PhysRevB.81.024102,yang2025ferroelectric,mahan2013many},
\begin{eqnarray}
\omega_{\rm TO}^2({\bf q})&=&(\Delta_0^2+v^2q^2)/m_{\rm op} \\
\omega_{\rm LO}^2({\bf q})&=&[\Delta_0^2+h_0^2+(v^2-w^2)q^2]/m_{\rm op}.
\end{eqnarray}
The dipole-dipole interaction therefore produces the characteristic LO-TO splitting in polar crystals. Since this splitting is typically large, the high-energy LO branch can be safely integrated out, and the low-lying TO mode dominates the ferroelectric instability. For a TO eigenmode with polarization vector ${\bf e}_{\alpha}$, one may parameterize the displacement field as~\cite{abrikosov2012methods,mahan2013many,kittel1963quantum}
\begin{equation}
{\bf u}^{\rm TO}({\bf r})
= \frac{1}{\sqrt{M_{\rm op}}} \sum_{\alpha=x,y,z}
\phi_{\alpha}({\bf r}) {\bf e}_{\alpha},
\end{equation}
where $\phi_{\alpha}({\bf r})$ denotes the field operator of the soft TO phonon.  Under this rescaling, the inertial part becomes 
\begin{equation}
\frac{1}{2}M_{\rm op}(\partial_tu^{\rm TO}_{\alpha})^2=\frac{1}{2}(\partial_t\phi_{\alpha})^2
\end{equation}
so that $\phi_\alpha$ is canonically normalized. 

When the electronic motion is confined to an interface plane (e.g., a 2DEG at $z=0$), momentum conservation requires that the relevant phonon wave vector ${\bf q}$ lies within the plane. Consequently, only the component polarized perpendicular to the interface (along the $z$ direction) couples efficiently to the 2DEG. Physically, this follows from the fact that the interfacial electric field generated by the dipolar lattice distortion is maximal for out-of-plane polarization, while in-plane transverse distortions do not produce a substantial longitudinal electric field acting on the confined electrons (See Sec.~\ref{seccou}).

Retaining only the $z$-polarized TO component and focusing on the static ground-state configuration, for which the inertial term vanishes, the above formulation naturally reduces to the effective polar-phonon Hamiltonian adopted in the main text.

\section{Coupling between electrons and polar phonons}
\label{seccou}

As mentioned in the main text, in a centrosymmetric bulk system the TO displacement does not couple linearly to the electronic density, and the leading electron–phonon coupling vanishes~\cite{PhysRevB.100.226501}. At an interface, however, inversion symmetry is explicitly broken. This symmetry reduction allows a finite linear coupling between the electrons and the TO mode. 

In the present study, the effective coupling strength is taken directly from experimentally extracted values of the electron-phonon coupling constant~\cite{norman2026superconductivity,Liu2023KTO_tunableSC}. Then, at fixed electron density, as assumed throughout this work, the analysis does not depend on a specific microscopic origin of the coupling, but instead treats it as a phenomenological parameter constrained by experiment. However, if one aims to investigate the carrier-density dependence of the ordered state, the microscopic origin of the coupling would become essential, since different mechanisms may lead to distinct doping evolutions of the effective interaction. For clarity and completeness, we briefly outline two representative microscopic mechanisms proposed in the literature that can give rise to such a linear interfacial coupling.

\subsection{Rashba spin-orbit-coupling-mediated mechanism}

At an interface lacking inversion symmetry, the electronic motion in the 2DEG is subject to Rashba spin-orbit coupling (SOC), arising from the interfacial electric field perpendicular to the plane~\cite{Wu2010SpinDynamics,Manchon2020Rashba,Bercioux2015RashbaReview}. The corresponding interaction reads~\cite{PhysRevB.93.235433,PhysRevB.92.155414,Bychkov1984Rashba}
\begin{equation}
V_{\rm R}(r) = \alpha_R({\bf r}) [\hat{\bf z}\times(-i{\bm \nabla})] \cdot {\boldsymbol {\hat \sigma}},
\end{equation}
where $\alpha_R$ is the Rashba coefficient and ${\boldsymbol {\hat \sigma}}$ denotes the Pauli matrices.

A TO displacement polarized along the $z$ direction modifies the local interfacial electric field and hence the Rashba coupling strength. To leading order, we expand
\begin{equation}
\alpha_R({\bf r})
= \alpha_R^{(0)} + \delta\lambda_R u_z({\bf r}),
\end{equation}
where $\delta\lambda_R$ characterizes the linear response of the Rashba coefficient to the lattice distortion~\cite{PhysRevB.105.224503,norman2026superconductivity,PhysRevResearch.5.023177,7l88-12m5}.

Using $u_z({\bf r}) =  \phi_z({\bf r})/\sqrt{M_{\rm op}}$, this expansion generates an effective electron-phonon interaction term, and adopting the symmetrized (Hermitian) form with average momentum yields
\begin{equation}
H_{\rm e\text{-}ph}^{\rm (R)}
=\frac{1}{\sqrt{M_{\rm op}}}
\sum_{{\bf k},{\bf q}}\sum_{ss'}
\delta\lambda_{R}
\,\phi_z({\bf q})
\,\psi^\dagger_{{\bf k+q}s}\Big\{
\Big[
\Big(\hat{\bf z}\times\frac{{\bf k}+{\bf k+q}}{2}\Big)
\Big]
\cdot {\boldsymbol {\hat \sigma}}\Big\}_{ss'}\,
\psi_{\bf ks'},
\end{equation}
which is linear in the TO displacement. Transforming to the helical basis that diagonalizes the Rashba Hamiltonian~\cite{Wu2010SpinDynamics} and retaining the intraband contribution, the electron-phonon coupling becomes
\begin{eqnarray}
H_{\rm e\text{-}ph}^{\rm (R)}&=&
\frac{1}{\sqrt{M_{\rm op}}}
\sum_{{\bf k}{\bf q}}
\delta\lambda_{R}
\phi_z({\bf q})
\sum_{\nu=\pm}\nu\Big|\frac{{{\bf k}+{\bf k+q}}}{2}\Big|\Big[\frac{1+e^{i(\theta_{\bf k}-\theta_{\bf k+q})}}{2}\Big]\psi^{\dagger}_{{\bf k+q},\nu}
\psi_{{\bf k},\nu}\nonumber\\
&=&\frac{1}{\sqrt{M_{\rm op}}}
\sum_{{\bf k}{\bf q}}\sum_{\nu=\pm}
{\nu|{\bf k}|}\,\delta\lambda_{R}
\phi_z({\bf q})\psi^{\dagger}_{{\bf k+q/2},\nu}
\psi_{{\bf k-q/2},\nu}\Big[\frac{1+e^{i(\theta_{\bf k_-}-\theta_{\bf k_+})}}{2}\Big]\nonumber\\
&=&\sum_{\bf kq}\sum_{\nu=\pm}g^{\nu}_{\bf k_+,k_-}\phi_z({\bf q})\psi^{\dagger}_{{\bf k+q/2},\nu}
\psi_{{\bf k-q/2},\nu},\quad\quad g^{\nu}_{\bf k_+,k_-}=\frac{{\nu|{\bf k}|}\,\delta\lambda_{R}}{\sqrt{M_{\rm op}}}\Big[\frac{1+e^{i(\theta_{\bf k_-}-\theta_{\bf k_+})}}{2}\Big],
\end{eqnarray}
with $\theta_{\bf k}$ being the angle of momentum ${\bf k}$. This coupling is symmetry allowed when inversion symmetry is broken at interface, and it directly links the out-of-plane polar lattice distortion to the in-plane electronic degrees of freedom, thereby providing a natural microscopic mechanism for the interfacial linear electron-phonon coupling.

\subsection{Bound-polarization-charges-mediated mechanism}

In addition to the Rashba-mediated mechanism discussed above, 
a linear electron-phonon coupling can also arise from the electrostatic 
potential generated by polarization bound charges associated with interfacial polar displacements which modify the electrostatic potential seen by the 2DEG~\cite{Hwang2012OxideInterfaces,Nakagawa2006PolarCatastrophe,Zubko2011InterfacePhysics,Garcia2010FerroelectricControl}. For an interfacial geometry where the TO displacement is primarily polarized 
perpendicular to the plane, the dominant contribution originates from 
the spatial variation of the polarization field along the $z$ direction, 
which produces an effective sheet-like bound charge near the interface. Specifically, in a polar crystal, a lattice displacement along $z$ induces a polarization field~\cite{yang2024first,cochran1981soft,cochran1961crystal,cowley1965theory,cochran1969dynamical,yelon1971neutron,tadmor2002polarization,tadmor2002polarization,zhong1994phase,PhysRevB.52.6301,PhysRevB.55.6161,PhysRevLett.84.5427,naumov2004unusual,PhysRevLett.97.157601,PhysRevLett.108.257601}
\begin{equation}
P_z({\bf r},z) =P_z({\bf r})f(z),\quad\quad   P_z({\bf r})=\frac{z^{*}u_z({\bf r})}{\Omega_{\rm cell}},  
\end{equation}
where $z^{*}$ is the Born effective charge~\cite{tadmor2002polarization,tadmor2002polarization,zhong1994phase,PhysRevB.52.6301,PhysRevB.55.6161,PhysRevLett.84.5427,naumov2004unusual,PhysRevLett.97.157601,PhysRevLett.108.257601}, $\Omega_{\rm cell}$ is the unit-cell volume, 
and $f(z)$ describes the interfacial envelope of the TO displacement  over a finite thickness~\cite{Stengel2006}.

The corresponding bound charge density is given by~\cite{jackson2012classical}
\begin{equation}
\rho_{\rm b}({\bf r},z)
= - \nabla \cdot {\bf P}({\bf r},z)=- \nabla_{\parallel} \cdot {\bf P}_{\parallel}({\bf r},z)-\partial_zP_z({\bf r},z)=-\partial_zP_z({\bf r},z).
\end{equation}
Here we have used that for the TO mode the in-plane polarization is transverse, so its longitudinal (divergent) component vanishes, i.e., $\nabla_{\parallel} \cdot {\bf P}_{\parallel}({\bf r},z)=0$.  After Fourier transformation in the in-plane coordinates, 
\begin{equation}
\rho_{\rm b}({\bf q},z)=-\partial_zP_z({\bf q},z)=-P_{z}({\bf q})\partial_zf(z).
\end{equation}
The induced electrostatic potential satisfies the Poisson equation~\cite{jackson2012classical}
\begin{equation}
[\partial_z^2-|{\bf q}|^2]\phi_b({\bf q},z) =-\rho_{\rm b}({\bf q},z)/[\varepsilon_0\varepsilon_{\rm eff}({\bf q})]
\end{equation}
where $\varepsilon_{\rm eff}({\bf q})$ accounts for effective dielectric screening  from the surrounding media. Solving for the potential at the interface ($z=0$) gives
\begin{equation}
\phi_b({\bf q},0)
=
\int dz'\,
\frac{\rho_{\rm b}({\bf q},z')\,e^{-|{\bf q}||z'|}}
{2|{\bf q}|\varepsilon_0\varepsilon_{\rm eff}({\bf q})}.
\end{equation}
Substituting $\rho_{\rm b}({\bf q},z)$ yields
\begin{equation}
\phi_b({\bf q},0)
=- P_z({\bf q})
\int dz'\,
\frac{\partial_{z'} f(z')\, e^{-|{\bf q}||z'|}}{
2|{\bf q}|\varepsilon_0\varepsilon_{\rm eff}({\bf q})}.
\end{equation}

The induced potential couples to the electronic density operator~\cite{kittel1963quantum,mahan2013many}
\begin{equation}
H_{\rm e\text{-}ph}=-e\int{d{\bf r}}\,\rho_e({\bf r})\phi_b({\bf r},z=0)=-e\int{d{\bf r}}\sum_{\bf q}\,\rho_e({\bf q})e^{i{\bf q}\cdot{\bf r}}\sum_{\bf q'}\phi_b({\bf q'},z=0)e^{i{\bf q'}\cdot{\bf r}}=\sum_{\bf q}\rho_e(-{\bf q})P_{z}({\bf q})F({\bf q})
\end{equation}
with
\begin{equation}
F({\bf q})=e\int{dz'}\partial_{z'}f(z')e^{-|{\bf q}||z'|}/[2|{\bf q}|\varepsilon_0\varepsilon_{\rm eff}({\bf q})]
\end{equation}
Using
\begin{equation}
P_z({\bf q})
=
\frac{z^*}{\sqrt{M_{\rm op}}\,\Omega_{\rm cell}}
\phi_z({\bf q}),
\end{equation}
we finally obtain
\begin{equation}
H_{\rm e\text{-}ph}
=
\sum_{\bf q}
g_{\bf q}\,
\rho_e(-{\bf q})\,
\phi_z({\bf q}),
\qquad
g_{\bf q}
=
\frac{z^* F({\bf q})}
{\sqrt{M_{\rm op}}\,\Omega_{\rm cell}}.
\end{equation}
where the density operator is given by $\rho_e(-{\bf q})=\sum_{\bf k\alpha}
\psi^{\dagger}_{{\bf k}+{\bf q/2},\alpha}
\psi_{{\bf k-q/2},\alpha}$~\cite{mahan2013many}. Unlike the Rashba-mediated coupling, the bound-charge mechanism couples directly to the total electronic density and is spin/band-independent at the vertex level.

It is worth emphasizing that the two mechanisms exhibit qualitatively different momentum dependences. At present, it remains unclear which mechanism dominates at the KTaO$_3$ interface, as relative strength of the two couplings depends sensitively on details such as the interfacial electric field, orbital composition of the conduction bands, dielectric screening, and the spatial profile of the electronic wave function across the interface. Therefore, in the present work we fix the carrier density and adopt experimentally extracted values for the effective coupling strength. This procedure avoids assumptions about the underlying microscopic mechanism while incorporating the interaction scale constrained by experiment.

\section{A general formulation model}

In this section, for completeness, we present a general formulation without assuming the electron–phonon coupling $g^{\alpha}_{{\bf k+{\bf Q},k}}$ to be momentum independent.  Specifically, starting from the electron-phonon coupled Hamiltonian in the main text, following the approach of Lee, Rice, and Anderson for 1D Fr\"ohlich-type coupling~\cite{lee1974conductivity}, generalized here to 2D, one can single out the polar phonon mode at ${\bf Q}$ and introduce a Landau-type order parameter~\cite{abrikosov2012methods} 
\begin{equation}
\Phi_{\bf Q} = -2\langle \phi_z({\bf Q}) \rangle ,~~~~\Phi^*_{\bf Q} =- 2\langle \phi_z^*({\bf Q}) \rangle =- 2\langle \phi_z(-{\bf Q}) \rangle.
\end{equation}
The sign convention in the definition of $\Phi_{\bf Q}$ here is essential. It is required to guarantee the correct energetic competition between the lattice distortion and the electronic reconstruction at the mean-field level, as it ensures that the condensation of the phonon mode lowers the electronic energy in the mean-field Hamiltonian and leads to a stable ordered state. Its microscopic origin and physical interpretation will be discussed in Sec.~\ref{secco}.

The resulting mean-field Hamiltonian is written as
\begin{eqnarray}
{\bar H}=\frac{1}{2}\sum_{\bf k,\alpha}\Psi^{\dagger}_{\bf k\alpha}\left(\begin{array}{cc}
\xi_{{\bf k_+},\alpha} & -g^{\alpha}_{{\bf k_+,k_-}}\Phi_{\bf Q} \\
-(g^{\alpha}_{{\bf k_+,k_-}}\Phi_{\bf Q})^*& \xi_{{\bf k_-},\alpha} 
\end{array}\right)\Psi_{\bf k\alpha}\!+\!\frac{2\omega^2_{\bf Q}|\Phi_{\bf Q}|^2}{8}\!+\!\frac{6b|\Phi_{\bf Q}|^4}{64},
\end{eqnarray}
with the spinor $\Psi^{\dagger}_{\bf k\alpha}=(\psi^{\dagger}_{{\bf k_+},\alpha},\psi^{\dagger}_{{\bf k_-},\alpha})$ and ${\bf k}_{\pm}={\bf k}\pm{\bf Q}/2$.

Diagonalizing this mean-field Hamiltonian~\cite{abrikosov2012methods,mahan2013many,schrieffer1964theory}, we
obtain the quasiparticle spectrum
\begin{equation}
E_{{\bf k}\alpha}^{\pm}=\frac{\xi_{\bf k_-,\alpha}+\xi_{\bf k_+,\alpha}}{2}\pm{\sqrt{\Big(\frac{\xi_{\bf k_-,\alpha}-\xi_{\bf k_+,\alpha}}{2}\Big)^2+|g^{\alpha}_{{\bf k_+,k_-}}\Phi_{\bf Q}|^2}},
\end{equation}
and the self-consistency mean-field gap equation
\begin{equation}
\frac{\omega^2_{\bf Q}\Phi_{\bf Q}}{2}+\frac{3b\Phi_{\bf Q}|\Phi_{\bf Q}|^2}{8}=\frac{1}{2}\sum_{{\bf k}\alpha}\frac{g^{\alpha}_{{\bf k_+,k_-}}\Phi_{\bf Q}[f(E_{\bf k\alpha}^-)-f(E_{\bf k\alpha}^+)]}{2\sqrt{[(\xi_{\bf k_-,\alpha}-\xi_{\bf k_+,\alpha})/2]^2+|g^{\alpha}_{{\bf k_+,k_-}}\Phi_{\bf Q}|^2}}.
\end{equation}
Consequently, One immediately sees that the gap opening is momentum and band dependent,
\begin{equation}
|\Delta_{\bf k}^{\alpha}|=|g^{\alpha}_{{\bf k_+,k_-}}\Phi_{\bf Q}|,
\end{equation}
which inherits the full structure of the electron–phonon vertex. The resulting quasiparticle spectrum therefore reflects both the Fermi-surface geometry and the band dependence of the interaction.

A closely analogous situation arises in BCS theory for realistic superconducting metals~\cite{schrieffer1964theory}, where the pairing interaction generally depends on momenta and band indices due to complex Fermi surfaces and multiband orbital structures. Although the interaction is momentum and band dependent in principle, it is often approximated by an effective constant within a narrow energy window around the Fermi surface, leading to the standard BCS formulation~\cite{schrieffer1964theory,abrikosov2012methods}. This simplification is well justified when the \emph{pairing symmetry is fixed}: renormalization-group (RG) analyses~\cite{RevModPhys.66.129,RevModPhys.84.299} show that microscopic interaction details can be absorbed into a renormalized effective coupling constant, while the form of the gap equation remains unchanged. 

In the present context, we follow the same controlled approximation: rather than resolving the full microscopic structure of the interaction vertex, we retain its effective low-energy projection in the relevant ordering channel. In this projected description, the microscopic momentum dependence of the vertex is absorbed into an effective coupling constant defined on the Fermi surface, yielding the simplified order parameter 
\begin{equation}
\Delta=g\Phi_{\bf Q},
\end{equation}
where the effective coupling $g$ represents the projected value of
$g^{\alpha}_{{\bf k+,k_-}}$ near the Fermi surface.

It should be emphasized that the main conclusions and  the central results presented in the present study are not tied to this simplification. Much like the relation between BCS and Eliashberg theory~\cite{RevModPhys.84.1383,RevModPhys.62.1027}, incorporating a more microscopic treatment (e.g., with full momentum/frequency dependence and self-energy effects) primarily leads to quantitative renormalizations of parameters and  quantitative corrections, while leaving the qualitative structure of the low-energy ordering instability unchanged. 
 We expect the same reasoning to apply to the instability of the composite order proposed in the present study.

\section{Electronic conductivity}

In this section, we derive the electronic conductivity, taking into account both 
the quasiparticle contribution and the sliding contribution associated with the collective mode. 
The total conductivity can be written schematically as
\begin{equation}
\sigma_{ij}=\sigma^{\rm q}_{ij}+\sigma^{\rm s}_{ij},
\end{equation}
where $\sigma^{\rm q}_{ij}$ arises from fermionic quasiparticles within the relaxation-time approximation~\cite{mahan2013many,kittel1963quantum}, 
while $\sigma^{\rm s}_{ij}$ originates from the sliding motion of the composite ordered state~\cite{gruner1985charge,gruner1988dynamics}.
We treat these two contributions separately in the following.

\subsection{Quasiparticle contribution}

This anisotropic Fermi-surface reconstruction manifests electronic nematicity in the quasiparticle excitations. Using the quasiparticle spectrum given in the main text,
\begin{equation}
E_{{\bf k}\alpha}^{\pm}=({\xi_{\bf k_-,\alpha}+\xi_{\bf k_+,\alpha}})/{2}\pm{\sqrt{[(\xi_{\bf k_-,\alpha}-\xi_{\bf k_+,\alpha})/2]^2+|\Delta|^2}},
\end{equation}
the corresponding dc conductivity can be evaluated within the relaxation-time approximation as~\cite{mahan2013many,kittel1963quantum}
\begin{equation}
\sigma^{q}_{ij}=e^2\tau\sum_{\bf k} v^-_i({\bf k})\,v^-_j({\bf k})\, f'\!\left(E_{\bf k}^-\right)+e^2\tau\sum_{\bf k} v^+_i({\bf k})\,v^+_j({\bf k})\, f'\!\left(E_{\bf k}^+\right),
\label{eq:sigma_q}
\end{equation}
where $i,j=x,y$. Here ${\bf v}^{\pm}({\bf k})=\nabla_{\bf k} E_{\bf k}^{\pm}$ is the quasiparticle group velocity and 
$f'(E)\equiv -\partial f(E)/\partial E$ is the energy derivative of the Fermi-Dirac distribution. The factor $f'(E)$ restricts the transport to states within an energy window of order $T$ around the Fermi level, and in the low-temperature regime one has
\begin{equation}
f'\!\left(E_{\bf k}^{\pm}\right)\xrightarrow{\mathrm{low}~~ T}\delta\!\left(E_{\bf k}^{\pm}\right).
\end{equation}
so that only quasiparticle states at the Fermi surface contribute to transport. Since $E_{\bf k}^+>0$ for all ${\bf k}$, the $+$ branch does not intersect the Fermi level and does not contribute the conductivity at low-temperature regime. Then, Eq.~(\ref{eq:sigma_q}) therefore reduces to a Fermi-surface  integral,
\begin{equation}
\sigma^{q}_{ij}=e^2\tau\int_{\rm FS}\frac{d\ell}{(2\pi)^2}\,
\frac{v^-_i({\bf k})\,v^-_j({\bf k})}{|{\bf v}^-({\bf k})|}.
\label{eq:sigma_q_FS}
\end{equation}
where $d\ell$ denotes the line element along the reconstructed Fermi contour.

Once $|\Delta|$ becomes finite, the reconstructed Fermi surface develops strong anisotropy (e.g., peanut-shaped), leading to a pronounced angular variation of the velocity components $v^-_x({\bf k})$ and $v^-_y({\bf k})$ along the Fermi contour. Consequently, the Fermi-surface averages of $(v^-_x)^2$ and $(v^-_y)^2$ become unequal, yielding $\sigma^{q}_{xx}\neq\sigma^{q}_{yy}$ and hence a finite nematic response in the quasiparticle transport channel. Importantly, within the relaxation-time approximation, the conductivity is determined solely by Fermi-surface averages of quadratic velocity components. The resulting nematic response is therefore strictly two-fold symmetric.  Specifically, the longitudinal quasiparticle conductivity \(\sigma^q_L(\theta)\), defined as the projection of the conductivity tensor along the direction of the applied electric field \(\hat{\bf e}_\theta=(\cos\theta,\sin\theta)\),  
$\sigma^q_L(\theta)
= \hat e_i \sigma^{q}_{ij} \hat e_j
= \sigma^{q}_{xx}\cos^2\theta
+ \sigma^{q}_{yy}\sin^2\theta$, exhibits a characteristic twofold anisotropy of the form
\begin{equation}
\sigma^q_L(\theta)
= [(\sigma^{q}_{xx}+\sigma^{q}_{yy})/2]  
+ [(\sigma^{q}_{xx}-\sigma^{q}_{yy})/2] \cos 2\theta. 
\label{eq:sigma_theta}
\end{equation} 
which contains only a $\cos 2\theta$ ($C_2$) angular dependence. No higher angular harmonics arise within this quasiparticle transport channel,  which differs qualitatively from the experimentally observed angular dependence, where higher harmonics beyond $C_2$ symmetry are clearly present~\cite{PhysRevX.15.021018,Zhang2023KTO111_rotSymBreak}.

\subsection{Sliding contribution}

In this part, we derive the sliding (collective) contribution to the conductivity arising from the phase dynamics of the density wave. We start from the electronic density-wave modulation introduced in the main text,
\begin{align}
\rho_e({\bf r})=2\chi|\Delta|
\cos\!\left({\bf Q}\cdot{\bf r}+\Theta\right),
\end{align}
where {\small $\chi={\omega^2_{\bf Q}}/(2g^2)+{b|\Delta|^2}/(8g^4)$}. Here $\Theta$ is the phase associated with translational symmetry breaking~\cite{gruner1985charge,gruner1988dynamics}, which determines the real-space shift of the modulation.

Using charge conservation,
\begin{equation}
\partial_t[-e\rho_e({\bf r})]+\nabla\cdot{\bf j}=0,
\end{equation}
and allowing for slow time-dependent phase fluctuations
\begin{equation}
\Theta({\bf r},t)=\Theta_0({\bf r})+\delta\Theta({\bf r},t),
\end{equation}
we obtain
\begin{equation}
\nabla\cdot{\bf j}=-2e\chi|\Delta|(\partial_t\delta\Theta)
\sin\!\left({\bf Q}\cdot{\bf r}+\Theta_0\right).
\end{equation}
Solving for the current consistent with the spatial modulation gives
\begin{equation}
{\bf j}=2e\chi|\Delta|(\partial_t\delta\Theta)\cos({\bf Q}\cdot{\bf r}+\Theta_0){\bf Q}/Q^2. 
\end{equation}

In realistic materials, the density-wave order is typically pinned by impurities~\cite{PhysRevB.17.535}, commensurability effects, or lattice imperfections~\cite{gruner1985charge,gruner1988dynamics}, which gives rise to an effective pinning potential~\cite{maki1986thermal,PhysRevB.17.535}  
\begin{equation}
V=-en_cQ^{-1}E_T\cos({\bf Q}\cdot{\bf r}+\Theta),
\end{equation}
where $E_T$ denotes the threshold electric field required to depin the density wave, and $n_c$ is the condensate (collective) carrier density associated with the density-wave order. As demonstrated by Fukuyama and Lee~\cite{PhysRevB.17.535}, when the impurity potential is either strong or the impurity concentration is dilute, the CDW
distorts to maximize the interaction with the impurity potential, resulting in the relation of the equilibrium-configuration spatial  average, 
\begin{equation}
\langle\cos({\bf Q}\cdot{\bf r}+\Theta_0)\rangle\approx1
\end{equation}
which corresponds to energy minimization with respect to the pinning potential. 

Upon spatial averaging, the collective current is given by 
\begin{equation}
{\bf j}=2e\chi|\Delta|\langle(\partial_t\delta\Theta)\rangle{\bf Q}/Q^2.
\end{equation}
Thus, the collective current is entirely determined by the phase velocity of the density wave.

In the presence of an external electric field ${\bf E}$, the coupling term contributes an additional driving energy $-e n_c Q^{-1}E_{\parallel}\Theta$~\cite{gruner1985charge,gruner1988dynamics,maki1986thermal}, where $E_{\parallel}={\bf E}\cdot\hat{\bf Q}$. The phase dynamics is taken to be relaxational~\cite{gruner1985charge,gruner1988dynamics},
\begin{equation}
n_c\gamma\partial_t\Theta=-{\partial_{\Theta}[V-en_cQ^{-1}E_{\parallel}\Theta]},
\end{equation}
which yields a driven, damped sine-Gordon equation~\cite{barone1982physics,drazin1989solitons,scott2003nonlinear} 
\begin{equation}
\gamma\partial_t\delta\Theta=\frac{2e}{\hbar}Q^{-1}E_{\parallel}-\frac{2e}{\hbar}Q^{-1}E_T[\sin(\delta\Theta)\langle\cos({\bf Q}\cdot{\bf r}+\Theta_0)\rangle+\cos(\delta\Theta)\langle\sin({\bf Q}\cdot{\bf r}+\Theta_0)\rangle]\approx\frac{2e}{\hbar}Q^{-1}E_{\parallel}-\frac{2e}{\hbar}Q^{-1}E_T\sin(\delta\Theta),
\end{equation}
 where $\gamma$ is a phenomenological damping coefficient.
 
In the overdamped limit, the driven sine-Gordon equation reduces to the dynamics of a phase in a tilted periodic potential, whose time-averaged velocity can be evaluated analytically over one period, yielding 
\begin{equation}
\langle(\partial_t\delta\Theta)\rangle={\rm sgn}(E_{\parallel})\frac{2e}{\hbar\gamma{Q}}\sqrt{E_{\parallel}^2-E_T^2}~\theta(|E_{\parallel}|-E_T).
\end{equation}
with $\theta(x)$ being the Heaviside step function. Consequently,  for $|E_{\parallel}|>E_T$, the phase enters a running state.  Substituting into the current expression gives the ideal sliding current
\begin{equation}
{\bf j}_s=\chi|\Delta|\frac{\bf Q}{Q^2}\frac{4e^2}{\hbar\gamma{Q}}{\rm sgn}(E_{\parallel})\sqrt{E_{\parallel}^2-E_T^2}~\theta(|E_{\parallel}|-E_T).
\end{equation}
This yields the characteristic square-root onset above the depinning threshold~\cite{gruner1985charge,gruner1988dynamics}.

In realistic disordered systems, spatially nonuniform pinning creates a landscape of metastable configurations separated by finite energy barriers~\cite{gruner1985charge,gruner1988dynamics}. Instead of a sharp depinning onset, the phase advances via thermally activated or quantum tunneling processes between adjacent minima. The ideal square-root onset is therefore replaced by a creep regime governed by the tunneling probability~\cite{PhysRevLett.42.1498,PhysRevLett.45.1978,PhysRevLett.51.1592}. In practice, experimental nonlinear transport in density-wave systems is often described by an empirical smooth creep form~\cite{gruner1985charge,PhysRevLett.42.1498,PhysRevLett.45.1978,PhysRevLett.51.1592}, 
\begin{equation}
\left\langle
\partial_t\delta\Theta
\right\rangle=\frac{2e\,{\rm sgn}(E_{\parallel})}{\hbar\gamma{Q}}
(|E_{\parallel}|-E_T)\exp\left(-\frac{E_0}{|E_{\parallel}|}\right),
\end{equation}
which yields~\cite{gruner1985charge,PhysRevLett.42.1498,PhysRevLett.45.1978,PhysRevLett.51.1592}
\begin{equation}
{\bf j}_s=\chi|\Delta|\frac{\bf Q}{Q^2}\frac{4e^2\,{\rm sgn}(E_{\parallel})}{\hbar\gamma{Q}}{(|E_{\parallel}|-E_T)}
\exp\left(-\frac{E_0}{|E_{\parallel}|}\right).
\end{equation}
Here $E_0$ characterizes the effective pinning barrier. Consequently, taking $x$ axis to be parallel to ${\bf Q}$, the corresponding sliding contribution to the field-dependent (secant) conductivity is given by 
\begin{equation}
\sigma_{ij}^s={4e^2\chi|\Delta|}/(\hbar{\gamma}Q^2)(1\!-\!E_T/|E_x|)e^{-E_0/|E_x|}\delta_{ix}\delta_{jx}.
\end{equation}
The exponential factor $\exp(-E_0/|E_\parallel|)$ here strongly suppresses the phase velocity at small electric fields and ensures that the collective current vanishes continuously as $E_\parallel<E_0$. Consequently, only sufficiently strong electric fields can generate an appreciable sliding contribution and allow it to compete with, or dominate over, the quasiparticle transport channel.\\

{\sl Discussions.---}The collective (sliding) current is constrained to flow along the density-wave ordering direction, as it originates from a uniform translation of the density modulation, and responds exclusively to the electric-field component $E_\parallel$.  This  current is therefore locked to ${\bf Q}$ rather than to the external electric field, i.e., in general, ${\bf j}_s$ is not parallel to ${\bf E}$ unless ${\bf E}\parallel{\bf Q}$. From the viewpoint of the electric-field direction, this implies that the sliding channel contributes to both longitudinal and transverse conductivity components, $\sigma_L^s(\theta)$ and $\sigma_T^s(\theta)$, where $\theta$ denotes the direction of the applied field.

It should be emphasized that this transverse response is \emph{not} a Hall effect but instead reflects the directional mismatch between the electric field and the density-wave ordering axis. Such transverse response is a generic feature of anisotropic systems~\cite{v12v-gl4n} and does not imply broken time-reversal symmetry or anomalous transport~\cite{PhysRev.37.405}.

The nonlinear transport therefore becomes strongly anisotropic, acquiring a pronounced nematic character tied directly to the density-wave orientation. For example, taking the $x$ axis parallel to ${\bf Q}$ and defining $\theta$ as the angle between the electric field and ${\bf Q}$, one finds that the longitudinal sliding conductivity $\sigma_L^s(\theta=0)$ is maximal while $\sigma_L^s(\theta=\pi/2)=0$. Meanwhile, the transverse component arises from the geometric projection of ${\bf j}_s$ onto the direction perpendicular to the field which vanishes at $\theta=0$ and $\theta=\pi/2$ and reaches its maximum near $\theta=\pi/4$.

Moreover, since $E_\parallel = {\bf E}\cdot{\bf Q} = E\cos\theta$, the nonlinear field dependence of the sliding current becomes intertwined with this angular projection. As a result, the field-angle dependence of both $\sigma_L^s(\theta)$ and $\sigma_T^s(\theta)$ is no longer purely sinusoidal, and higher angular harmonics are naturally generated. This provides a simple and intrinsic mechanism for the emergence of higher-order 
harmonics in transport nematicity. In contrast, the quasiparticle contribution, 
which arises from Fermi-surface anisotropy within linear response, remains 
purely twofold and does not generate higher angular harmonics. Moreover, its 
magnitude is parametrically small for the reconstructed Fermi surface 
considered here and cannot account for the experimentally observed strength 
of the transport nematicity~\cite{PhysRevX.15.021018,Zhang2023KTO111_rotSymBreak}.

\section{Additional discussions}

In this section, we present additional discussions.

\subsection{Composite order}
\label{secco}

The composite order parameter introduced in the main text is defined as
\begin{equation}\label{OP}
\Delta = -2 g\, \langle \phi_z({\bf Q}) \rangle, \quad  \Delta^* =- 2 g\, \langle \phi_z(-{\bf Q}) \rangle,
\end{equation}
The minus sign here is essential as it ensures a non-trivial solution of the self-consistent gap equation. 
Its physical origin is more fundamental: it reflects the relative phase that minimizes the electron-polarization coupling energy. 

Specifically, using the definition of the order parameter, the induced polarization density wave and electronic density wave are obtained as 
\begin{eqnarray}
P_z({\bf r})=-{z^*}/{(g\sqrt{M}\Omega_{\rm cell})}|\Delta|\cos({\bf Q}\cdot{\bf r}+\Theta).
\end{eqnarray}
and 
\begin{align}
\rho_e({\bf r})=2\chi|\Delta|
\cos\!\left({\bf Q}\cdot{\bf r}+\Theta\right),
\end{align}
Then, one immediately sees that for $g>0$, the polarization density wave and the electronic density modulation are out of phase. For $g<0$, they are in phase. Substituting these expressions into the coupling energy yields
\begin{equation}
F_{\rm int}\sim g\rho_eP_z\sim -2\chi|\Delta|^2\cos^2\!\left({\bf Q}\cdot{\bf r}+\Theta\right),
\end{equation}
which is strictly negative, independent of the sign of $g$.
The system therefore spontaneously selects the relative phase between polarization and charge modulation such that the coupling energy is minimized. 

Therefore, the minus sign in the definition of $\Delta$, which  allows a non-trivial ordered solution of the gap equation,  incorporate the energy-minimizing relative phase at the mean-field level, i.e., it encodes an energetically favorable composite phase-locking configuration.  Consequently, while the lattice alone remains in a quantum paraelectric phase and the electronic system alone does not spontaneously develop stripe order, their mutual coupling qualitatively alters the stability of the uniform state. The coupled electron-lattice system then lowers its free energy through the formation of a composite ordered state, in which electronic modulation and lattice polarization fluctuations develop self-consistently.

\subsection{Polarization density waves}

As discussed above, the composite order identified in this work arises from a cooperative electron–lattice instability at the near-ferroelectric KTaO$_3$ interface. Our results show that the uniform paraelectric state becomes unstable below a critical temperature once it is coupled to itinerant electrons. The resulting ordered state that involves intertwined electronic and lattice degrees of freedom cannot be reduced to a purely electronic or purely lattice-driven mechanism.

In this part, we also discuss the possible relevance of polarization density waves (PDWs), recently proposed in bulk SrTiO$_3$ from a purely lattice-driven mechanism~\cite{PhysRevLett.131.046801,Orenstein2025,PhysRevB.106.L140301}, and clarify their distinction from the present composite-order scenario. 

Recent theoretical work~\cite{PhysRevLett.131.046801} has proposed that flexoelectric coupling between TO and TA phonon modes in SrTiO$_3$ can induce a softening of the TA branch at finite momentum in the low-temperature limit. Such finite-$q$
 softening of the TA branch deviates from conventional Debye behavior of the acoustic phonons and becomes increasingly pronounced as the TO mode softens upon cooling to low-temperature limit.  When the flexoelectric coupling is sufficiently strong, this mechanism may drive a modulated lattice instability at low temperatures, giving rise to a PDW  characterized by spatially varying polarization~\cite{PhysRevLett.131.046801}.

Experimentally, deviations of the TA mode from Debye behavior around $T\approx20~$K in bulk SrTiO$_3$~\cite{Orenstein2025,PhysRevB.106.L140301}, while being weak in magnitude, have recently been interpreted as possible  signatures of a PDW tendency~\cite{Orenstein2025}.  However, several measurements performed at lower temperatures did not observe a further or systematic enhancement of the TA softening upon cooling~\cite{PhysRevLett.124.145901,Yamada1969JPSJ}, nor compelling evidence for a stabilized PDW phase~\cite{zhang2025nanoscale}. Meanwhile, recent studies in disordered systems suggest that disorder alone can generate non-Debye acoustic responses~\cite{Ding2025} without necessarily involving/driving a genuine modulated instability.   More directly, recent transmission electron microscopy (TEM) studies in bulk SrTiO$_3$~\cite{zhang2025nanoscale} indicate that below the cubic-to-tetragonal antiferrodistortive (AFD) transition at $T_{\rm AFD}\approx105~$K, short-range polar domains grow and organize into quasi-periodic structures extending over tens of nanometers. Yet upon further cooling into the quantum paraelectric regime below $\sim40~$K, this tendency reverses: the periodically arranged polar nanodomains fragment into smaller clusters, and short-range polar correlations weaken and eventually disappear at the lowest temperatures~\cite{zhang2025nanoscale}. Such behavior appears inconsistent with the stabilization of a robust long-wavelength modulated polarization state in the low-temperature limit.  One possible explanation is that the effective flexoelectric coupling may not be sufficiently strong to overcome competing interactions and stabilize a true PDW ground state in SrTiO$_3$. In addition, the instability may require mediation by fluctuations of the AFD order rather then the coupling by TA mode: such a mechanism would naturally be strongest near the AFD transition at $T_{\rm AFD}\approx105$~K, where AFD fluctuations are enhanced, but would weaken upon further cooling as the AFD order parameter hardens below $\sim 40$~K.

Importantly, it should be emphasized that the relevance of such a PDW scenario to the present work is limited. In KTaO$_3$, comparable non-Debye anomalies (finite-$q$ softening) of the TA mode have rarely been observed unambiguously in experiments~\cite{Farhi2000,He2025,PhysRevB.39.8666}. Moreover, unlike SrTiO$_3$, KTaO$_3$ does not undergo an AFD structural transition, and remains as cubic down
to low-temperature limit~\cite{PhysRevB.53.176}. The absence of this additional structural instability further reduces the plausibility of a lattice-driven finite-$q$ polarization instability in KTaO$_3$. For these reasons, an indirect mechanism mediated by PDW formation, where flexoelectric interactions first generate a modulated polarization state that subsequently induces electronic stripe order, appears unlikely in the present context. It is therefore more appropriate to directly examine the possibility of a composite-order instability at the KTaO$_3$ interface, as carried out in the present study.

\subsection{Interplay with ferroelectric ordering and superconductivity}

In the present work, we focus on a near-ferroelectric paraelectric interface, where the lattice remains close to instability but does not develop long-range ferroelectric order. By contrast, a near-paraelectric ferroelectric interface with established ferroelectric ordering would introduce additional symmetry breaking and static polarization fields. In such a regime, the resulting interplay with electronic degrees of freedom is expected to be qualitatively different. In particular, the onset of ferroelectric ordering typically hardens the soft polar mode~\cite{yang2025ferroelectric,yang2024first,cochran1981soft,cochran1961crystal,cowley1965theory,cochran1969dynamical,yelon1971neutron,tadmor2002polarization,tadmor2002polarization,zhong1994phase,PhysRevB.52.6301,PhysRevB.55.6161,PhysRevLett.84.5427,naumov2004unusual,PhysRevLett.97.157601,PhysRevLett.108.257601}, which may suppress the tendency toward composite-order formation by reducing the strength of critical lattice fluctuations. On the other hand, the emergence of a static polarization further enhances interfacial symmetry breaking and can modify the electronic structure and coupling channels~\cite{Krizman2024AdvMater}. The net effect is therefore nontrivial and cannot be inferred straightforwardly from the paraelectric limit. A systematic analysis of such a regime lies beyond the scope of the present study and remains an interesting direction for future work.

It is of considerable interest to explore how the composite electron-lattice order identified here may coexist or compete with superconductivity at the KTaO$_3$ interface. Since both instabilities originate from the same low-energy electronic states coupled to critical lattice fluctuations, their interplay is expected to be highly nontrivial at the microscopic level, potentially involving multi-order competition and mutual feedback effects. Whether nematicity and superconductivity are mutually exclusive or can coexist in an intertwined phase, possibly through a sequential symmetry-breaking transition, remains an open question. A detailed investigation of this issue lies beyond the scope of the present study and will be presented separately.

\section{Simulation parameters}

As mentioned in the main text, without loss of generality and in order to make the proposed mechanism more transparent, we approximate the electronic dispersion by isotropic parabolic bands, $\xi_{{\bf k},\alpha}=\hbar^2{k^2}/(2m)-\mu$, and hence, the symmetry breaking associated with the composite order reduces the in-plane rotational symmetry from continuous isotropy to $C_2$.

\begin{figure}[H]
\centering
  \includegraphics[width=5.8cm]{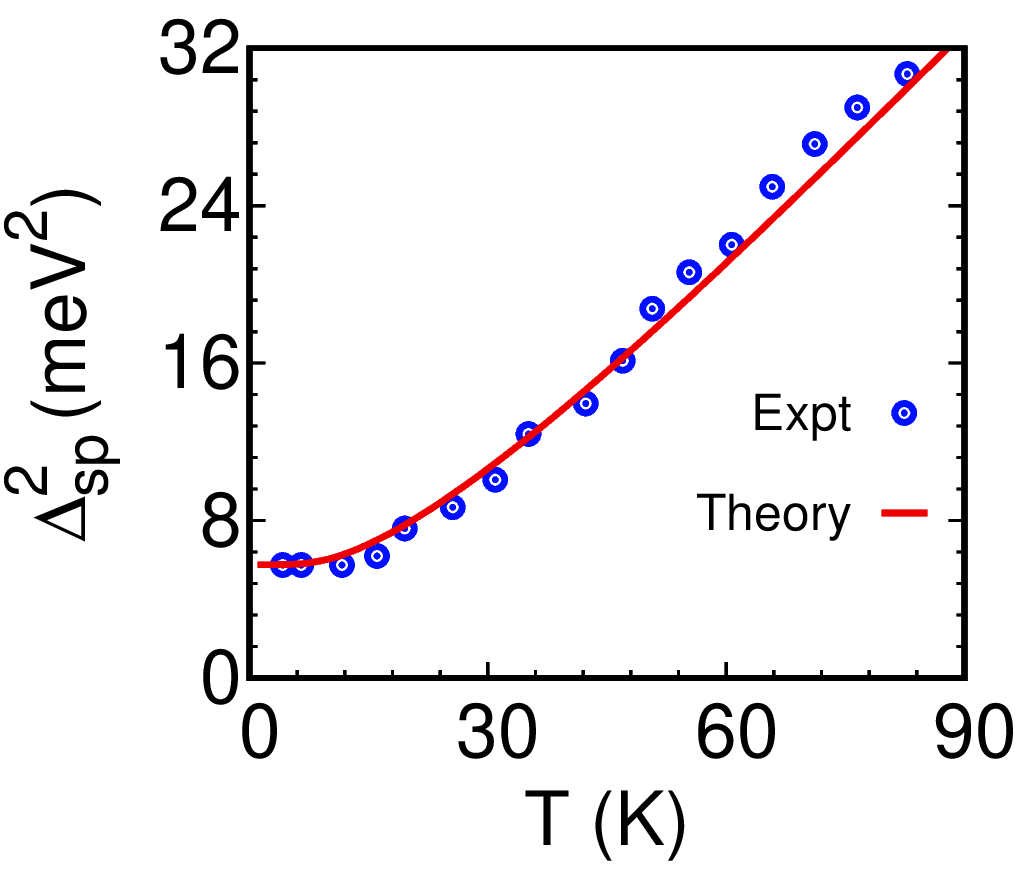}
\caption{Temperature dependence of the soft TO phonon gap in KTaO$_3$, calculated within the self-consistent phonon renormalization framework and compared with hyper-Raman scattering data from Ref.~\cite{PhysRevB.51.8046}.}
  \label{Sfigyc}
\end{figure}

For model parameters of polar phonons, we re-write  the harmonic restoring coefficien $a=\Delta_{\rm sp}^2$ with $\Delta_{\rm sp}$ being the TO-mode gap. According to hyper-Raman scattering measurements on KTaO$_3$, the low-temperature limit of the TO-mode gap is $\Delta_{\rm sp}(T\sim0)\approx2.4~$meV~\cite{PhysRevB.51.8046,PhysRev.157.396}, while the corresponding phonon velocity is $v=57~$\AA/ps~\cite{Rowley2014FE_QCP,PhysRevB.51.8046,PhysRev.157.396}.  In the present work, the phonon field is in fact defined as an interfacial effective field obtained after integrating over the finite confinement length along the $z$  direction and rescaling to preserve canonical normalization. Accordingly,  the quartic anharmonic coefficient $b$ should be understood as an effective interfacial parameter.  Within the self-consistent phonon renormalization framework~\cite{yang2024first,74d5-4hsw,Rowley2014FE_QCP}, the quadratic coefficient becomes temperature dependent and satisfies $a(T)=a(T=0)+3bC(T)$, where the fluctuation contribution is given by $C(T)=\sum_{\bf q}2n(\hbar\omega_q)/[\hbar\omega_q(T)]$~\cite{yang2024first,74d5-4hsw,Rowley2014FE_QCP} with $n(x)$ being the Bose-Einstein distribution function, and the renormalized phonon dispersion reads $\omega_q(T)=\sqrt{a(T)+v^2q^2}$.  The resulting temperature dependence of the TO-mode gap, $\Delta_{\rm sp}(T)=\sqrt{a(T)}$, is required to be consistent with the experimentally observed bulk evolution. By enforcing this consistency self-consistently [as shown in Fig.~\ref{Sfigyc}], one can determine the effective interfacial anharmonic coefficient $b$.

The low-energy sector of 2DEG formed at the KTaO$_3$(111) interface consists of three nearly degenerate conduction bands derived from the $t_{2g}$ manifold~\cite{norman2026superconductivity,Liu2023KTO_tunableSC}. Using the 2D density of states per spin~\cite{kittel1963quantum},
$N(0)=m/(2\pi\hbar^2)$,
the dimensionless electron-phonon coupling constant (BCS coupling constant~\cite{schrieffer1964theory,mahan2013many,abrikosov2012methods,RevModPhys.84.1383,RevModPhys.62.1027}) is 
\begin{equation}\label{BCSC}
\lambda={N(0)g^2}/\langle{\omega^2_{\bf k_F-k_F'}}\rangle_{\rm FS}.
\end{equation}
We adopted an experimentally estimated electron–phonon coupling constant $\lambda\approx 0.26$~\cite{norman2026superconductivity,Liu2023KTO_tunableSC}, an effective mass $m\approx 0.34 m_e$~\cite{Zou2015LTO_KTO_2DEG,PhysRevLett.108.117602}, and a carrier density $n_e = 6\times10^{13}\mathrm{cm^{-2}}$~\cite{Hua2024,PhysRevX.15.021018,Liu2021KTaO3_111_SC,Liu2023KTO_tunableSC} at the KTaO$_3$(111) interface. Then, the coupling strength $g$ is determined by Eq.~(\ref{BCSC}).

In the numerical implementation of the gap equation, the ordering wavevector $|{\bf Q}|$ is treated as a variational parameter at each temperature. 
For a given $T$, we evaluate $|\Delta({\bf Q})|$ over a range of $|{\bf Q}|$ values around the physically relevant momentum scale (guided by the analytical condition discussed in the main text) and identify the optimal wavevector that maximizes $|\Delta({\bf Q})|$, equivalently minimizing the free energy. 
The resulting optimal $|{\bf Q}|$ is then substituted back into the gap equation to obtain the self-consistent temperature-dependent gap magnitude $|\Delta(T)|$.

\begin{table}[htb]
\centering
  \caption{Specific parameters used in our numerical calculation. $m_e$ denotes the free electron mass.}  
\renewcommand\arraystretch{1.5}   \label{Table}
\begin{tabular}{l l l l l l l l}
    \hline
    \hline
    model parameter for electrons&\quad\quad\quad~$m$&\quad\quad\quad~$n_e$\\  
    &\quad\quad\quad$0.34m_e$&\quad\quad $6\times10^{13}/$cm$^2$\\
     \hline
    model parameters for polar phonons&\quad\quad\quad$\Delta_{\rm sp}$&\quad\quad\quad\quad\quad$b$&\quad\quad$v$\\   
    &\quad\quad2.4~meV&\quad\quad$3\times10^4~\text{meV}^3$\AA$^2$&\quad\quad$57~$\AA/ps\\
         \hline
    coupling constant&\quad\quad\quad$\lambda$\\   
    &\quad\quad\quad0.26\\
    \hline
    \hline
\end{tabular}\\
\end{table}

All conductivities shown in Fig.~2(a) and (c) of the main text are normalized by the angularly averaged value $\bar \sigma=(\sigma^q_{xx}+\sigma^q_{yy})/2$, in order to highlight the intrinsic anisotropy induced by the composite-order formation rather than the overall magnitude of the conductivity. In evaluating the sliding contribution to the conductivity [Fig.~2(c) and (d)], the in-plane electric field is decomposed as $E_x=E\cos\theta$
 where $\theta$ denotes the angle between the applied field and the ordering wave vector ${\bf Q}$.  Following Ref.~\cite{gruner1985charge}, we approximately take the characteristic pinning barrier as $E_0 = 5E_T$, where $E_T$ is the threshold field. In the calculation of sliding contribution, we focus on the strong-field (or weak-pining) regime and set $E=100E_T$,  such that the collective sliding mode is fully developed and the nonlinear contribution dominates over the pinned response. In this limit, the conductivity anisotropy arises predominantly from the angular modulation of the sliding channel. This choice ensures that the qualitative features shown in Fig.~2(c) and (d) are robust as a nearly field-independent form $\sigma_{ij}^s\approx{4e^2\chi|\Delta|}/({\gamma}Q^2)\delta_{ix}\delta_{jx}$.  We fix the damping rate $\gamma$ by requiring the sliding prefactor at $T=0$ to reproduce a fraction of the averaged conductivity,
\begin{equation}
\frac{4e^2\,\chi(0)\,|\Delta(0)|}{\hbar\,\gamma\,Q^2}=0.55\,\bar{\sigma},
\end{equation}
such that the resulting total conductivity yields a nematicity ratio consistent with experimental observations~\cite{PhysRevX.15.021018}. Then, its temperature dependence is evaluated by 
\begin{equation}
\sigma_{ij}^s(T)=\sigma_{ij}^s(T=0)\frac{\chi(T)|\Delta(T)|}{\chi(0)|\Delta(0)|}.
\end{equation}

Tuning the system closer to or further away from the ferroelectric critical point, via strain, pressure, or chemical substitution, as illustrated in Fig.~3(b), is modeled phenomenologically by a linear softening of the Landau coefficient, $a(x)=a(0)(1-x/x_{\rm cp})$~\cite{yang2024first,yang2025ferroelectric,Rowley2014FE_QCP}, in order to capture the enhancement of $T_c$ as the system approaches the ferroelectric critical point. The resulting $T_c$ enhancement should be regarded as an idealized estimate. In particular, in the immediate vicinity of the critical point, additional effects beyond homogeneous soft-mode softening may become important. Spatial inhomogeneity, the formation of polar nano-regions, disorder-induced fluctuations, and impurity scattering can all modify the effective critical behavior and potentially suppress or smear the proposed composite order. Therefore, the linear softening picture provides a transparent upper-bound scenario for the critical enhancement of $T_c$, while real materials may exhibit deviations due to mesoscopic and disorder-driven effects.

Furthermore, for $x < 0.96 x_{\rm cp}$, the calculated transition temperature always satisfies $k_B T_c(x) < \Delta_{\rm sp}(T=0,x)$. In this regime, the soft-mode gap remains larger than the thermal energy even at the transition temperature $T_c$ of the composite ordered state. As a consequence, the thermal renormalization of the soft mode, manifested through the temperature dependence of $a(T,x)$, is quantitatively weak and does not qualitatively modify the mean-field instability. Fluctuation-induced corrections to $T_c$ or to the self-consistent gap equation are therefore expected to be subleading. The mean-field estimate thus provides a controlled description, except in the asymptotic vicinity of the critical point where the soft-mode gap becomes parametrically small.


%